\documentclass[prl,twocolumn,showpacs,superscriptaddress,preprintnumbers,floatfix]{revtex4}

\usepackage{ifpdf}
\usepackage{dcolumn}
\usepackage{url}
\usepackage{amsmath}
\usepackage{amscd}
\usepackage{amsfonts}
\usepackage{amssymb}
\usepackage{bm}   
\usepackage{bbm}   
\usepackage{verbatim}
\usepackage{stmaryrd}
\usepackage{amsthm}

  \usepackage{graphicx}

\usepackage{color}

\theoremstyle{plain}    
\theoremstyle{plain}    
\theoremstyle{plain} 	\newtheorem{Cor}{Corollary}
\theoremstyle{plain} 	\newtheorem*{ProCor}{Proof}
\theoremstyle{plain} 	\newtheorem{The}{Theorem}
\theoremstyle{plain} 	\newtheorem*{ProThe}{Proof}
\theoremstyle{plain} 	\newtheorem{Prop}{Proposition}
\theoremstyle{plain} 	\newtheorem*{ProProp}{Proof}
\theoremstyle{plain} 	
\theoremstyle{plain}	
\theoremstyle{plain}	\newtheorem*{Def}{Definition} 
\theoremstyle{plain}	

\newcommand{\figref}[1]{Fig.~\ref{#1}}
\newcommand{\tableref}[1]{Table~\ref{#1}}
\newcommand{\refcite}[1]{Ref.~\cite{#1}}

\newcommand{\eM}     {\mbox{$\epsilon$-machine}}
\newcommand{\eMs}    {\mbox{$\epsilon$-machines}}
\newcommand{\EM}     {\mbox{$\epsilon$-Machine}}
\newcommand{\EMs}    {\mbox{$\epsilon$-Machines}}

\newcommand{\Process}{\mathcal{P}}

\newcommand{\MeasSymbol}   { {X} }
\newcommand{\meassymbol}   { {x} }
\newcommand{\BiInfinity}	{ \overleftrightarrow {\MeasSymbol} }
\newcommand{\biinfinity}	{ \overleftrightarrow {\meassymbol} }
\newcommand{\Past}	{ \overleftarrow {\MeasSymbol} }
\newcommand{\past}	{ {\overleftarrow {\meassymbol}} }
\newcommand{\pastprime}	{ {\past}^{\prime}}
\newcommand{\Future}	{ \overrightarrow{\MeasSymbol} }
\newcommand{\future}	{ \overrightarrow{\meassymbol} }
\newcommand{\futureprime}	{ {\future}^{\prime}}

\newcommand{\AllPasts}	{ { \overleftarrow {\rm {\bf \MeasSymbol}} } }

\newcommand{\CausalState}	{ \mathcal{S} }

\newcommand{\CausalStateSet}	{ \boldsymbol{\CausalState} }
\newcommand{\AlternateState}	{ \mathcal{R} }
\newcommand{\AlternateStatePrime}	{ {\cal R}^{\prime} }
\newcommand{\alternatestate}	{ \rho }
\newcommand{\alternatestateprime}	{ {\rho^{\prime}} }
\newcommand{\AlternateStateSet}	{ \boldsymbol{\AlternateState} }
\newcommand{\PrescientState}	{ \widehat{\AlternateState} }

\newcommand{\PrescientStateSet}	{ \boldsymbol{\PrescientState}}

\newcommand{\Prob}      {\Pr} 

\newcommand{\Cmu}		{C_\mu}
\newcommand{\hmu}		{h_\mu}
\newcommand{\EE}		{{\bf E}}

\newcommand{\PC}		{\chi}
\newcommand{\FuturePC}		{\PC^+}
\newcommand{\PastPC}		{\PC^-}
\newcommand{\CI}		{\Xi}
\newcommand{\ReverseMap}	{r}
\newcommand{\ForwardMap}	{f}

\newcommand{\forward}{+}
\newcommand{\reverse}{-}
\newcommand{\forwardreverse}{\pm} 
\newcommand{\FutureProcess}	{ {\Process}^{\forward} }
\newcommand{\PastProcess}	{ {\Process}^{\reverse} }
\newcommand{\FutureCausalState}	{ {\CausalState}^{\forward} }
\newcommand{\futurecausalstate}	{ \sigma^{\forward} }

\newcommand{\PastCausalState}	{ {\CausalState}^{\reverse} }
\newcommand{\pastcausalstate}	{ \sigma^{\reverse} }
\newcommand{\BiCausalState}		{ {\CausalState}^{\forwardreverse} }

\newcommand{\FutureCausalStateSet}	{ {\CausalStateSet}^{\forward} }
\newcommand{\PastCausalStateSet}	{ {\CausalStateSet}^{\reverse} }
\newcommand{\BiCausalStateSet}	{ {\CausalStateSet}^{\forwardreverse} }
\newcommand{\eMachine}	{ M }
\newcommand{\FutureEM}	{ {\eMachine}^{\forward} }
\newcommand{\PastEM}	{ {\eMachine}^{\reverse} }
\newcommand{\BiEM}		{ {\eMachine}^{\forwardreverse} }
\newcommand{\BiEquiv}	{ {\sim}^{\forwardreverse} }
\newcommand{\Futurehmu}	{ h_\mu^{\forward} }
\newcommand{\Pasthmu}	{ h_\mu^{\reverse} }
\newcommand{\FutureCmu}	{ C_\mu^{\forward} }
\newcommand{\PastCmu}	{ C_\mu^{\reverse} }
\newcommand{\BiCmu}		{ C_\mu^{\forwardreverse} }
\newcommand{\FutureEps}	{ \epsilon^{\forward} }
\newcommand{\PastEps}	{ \epsilon^{\reverse} }
\newcommand{\BiEps}	{ \epsilon^{\forwardreverse} }

\newcommand{\PastSim}	{ \sim^{\reverse} }

\newcommand{\TR}{\mathcal{T}}
\newcommand{\MSP}{\mathcal{U}}
\newcommand{\one}{\mathbf{1}}

\newif\ifpm 
\edef\tempa{\forwardreverse}
\edef\tempb{\pm}
\ifx\tempa\tempb
   \pmfalse
\else
   \pmtrue  
\fi

\def\figscale{.7}

\begin{document}

\title{Prediction, Retrodiction, and\\
The Amount of Information Stored in the Present}

\author{Christopher J. Ellison}
\email{cellison@cse.ucdavis.edu}
\affiliation{Complexity Sciences Center and Physics Department,
University of California at Davis, One Shields Avenue, Davis, CA 95616}

\author{John R. Mahoney}
\email{jrmahoney@ucdavis.edu}
\affiliation{Complexity Sciences Center and Physics Department,
University of California at Davis, One Shields Avenue, Davis, CA 95616}

\author{James P. Crutchfield}
\email{chaos@cse.ucdavis.edu}
\affiliation{Complexity Sciences Center and Physics Department,
University of California at Davis, One Shields Avenue, Davis, CA 95616}
\affiliation{Santa Fe Institute, 1399 Hyde Park Road, Santa Fe, NM 87501}

\date{May 19, 2009}

\bibliographystyle{unsrt}

\begin{abstract}
We introduce an ambidextrous view of stochastic dynamical systems, comparing
their forward-time and reverse-time representations and then integrating them
into a single time-symmetric representation. The perspective is useful
theoretically, computationally, and conceptually. Mathematically, we prove that
the excess entropy---a familiar measure of organization in complex systems---is
the mutual information not only between the past and future, but also between
the predictive and retrodictive causal states. Practically, we exploit the
connection between prediction and retrodiction to directly calculate the
excess entropy. Conceptually, these lead one to discover new system invariants
for stochastic dynamical systems: crypticity (information accessibility) and
causal irreversibility. Ultimately, we introduce a time-symmetric
representation that unifies all these quantities, compressing the two
directional representations into one. The resulting compression offers a
new conception of the amount of information stored in the present.
\end{abstract}

\pacs{
02.50.-r  
89.70.+c  
05.45.Tp  
02.50.Ey  
}
\preprint{Santa Fe Institute Working Paper 09-05-XXX}
\preprint{arxiv.org:0905.3587 [cond-mat.stat-mech]}

\maketitle




\section{Introduction}

``Predicting time series'' encapsulates two notions of directionality.
\emph{Pre}diction---making a claim about the future based on the past---is
directional. \emph{Time} evokes images of rivers, clocks, and actions in
progress. Curiously, though, when one writes a time series as a lattice of
random variables, any necessary dependence on time's inherent direction is
removed; at best it becomes convention. When we analyze a stochastic process to
determine its correlation function, block entropy, entropy rate, and the like,
we already have shed our commitment to the idea of \emph{forward} by virtue of
the fact that these quantities are defined independently of any perceived
direction of the process.

Here we explore this ambivalence. In making it explicit, we consider not only
predictive models, but also retrodictive models. We then demonstrate that it
is possible to unify these two viewpoints and, in doing so, we discover
several new properties of stationary stochastic dynamical systems. Along the
way, we also rediscover, and recast, old ones.

We first review minimal causal representations of stochastic processes, as
developed by \emph{computational mechanics}~\cite{Crut88a,Crut98d}. We extend
its (implied) forward-time representation to reverse-time. Then, we prove that
the mutual information between a process's past and future---the
\emph{excess entropy}---is the mutual information between its
forward- and reverse-time representations.

Excess entropy, and related mutual information quantities, are widely used
diagnostics for complex systems. They have been applied to detect the
presence of organization in dynamical systems~%
\cite{Fras90b,Casd91a,Spro03a,Kant06a}, in spin systems~%
\cite{Arno96,Crut97a,Feld98b}, in neurobiological systems~%
\cite{Tono94a,Bial00a},
and even in language, to mention only a few applications. For example, in
natural language the excess entropy ($\EE$) diverges with the number of
characters $L$ as $\EE \propto L^{1/2}$. The claim is that this reflects the
long-range and strongly nonergodic organization necessary for human
communication~\cite{Ebel94c,Debo08a}.

The net result is a unified view of information processing in
stochastic processes. For the first time, we give an explicit
relationship between the internal (causal) state information---the statistical
complexity~\cite{Crut88a}---and the observed information---the excess entropy.
Another consequence is that the forward and reverse representations are two
projections of a unified time-symmetric representation. From the latter it
becomes clear there are important system invariants that control how accessible
internal state information is and how irreversible a process is. Moreover, the
methods are sufficiently constructive that one can calculate the excess
entropy in closed-form for finite-memory processes.

Before embarking, we refer the reader to \refcite{Crut08c} for complementary
results, that we do not cover here, on the measure-theoretic relationships
between the above information quantities. The announcement of those results
and those in the present work appeared in \refcite{Crut08a}. Here we lay out
the theory in detail, giving step-by-step proofs of the main results and the
calculational methods.

\section{Optimal Causal Models}

Our approach starts with a simple analogy. Any process $\Pr(\Past,\Future)$ is
also a \emph{communication channel} with a specified input distribution
$\Pr(\Past)$~\footnote{Throughout, we follow the notation and definitions 
of Refs.~\cite{Cove06a,Crut98d}. In addition, when we say
$\protect\Future$, for example, this should be interpreted
as a shorthand for using $\protect\Future^L$ and then
taking an appropriate limit,
such as $\lim_{L \rightarrow \infty}$ or $\lim_{L \rightarrow \infty} 1/L$.}:
It transmits information from the \emph{past}
$\Past = \ldots \MeasSymbol_{-3} \MeasSymbol_{-2} \MeasSymbol_{-1}$ to the
\emph{future} $\Future = \MeasSymbol_0 \MeasSymbol_1 \MeasSymbol_2 \ldots$
by storing it in the present. $\MeasSymbol_t$ is the random variable for
the measurement outcome at time $t$. Our goal is also simply stated: We wish to
predict the future using information from the past. At root, a prediction is
probabilistic, specified by a distribution of possible futures $\Future$
given a particular past $\past$: $\Pr(\Future|\past)$. At a minimum, a
good predictor needs to capture \emph{all} of the information $I$ shared
between past and future: $\EE = I[\Past;\Future]$---the process's
\emph{excess entropy}~\cite[and references therein]{Crut01a}.

Consider now the goal of modeling---building a representation that allows not
only good prediction but also expresses the mechanisms producing a system's
behavior. To build a model of a structured process (a memoryful channel),
computational mechanics~\cite{Crut88a} introduced an equivalence relation
$\past \sim \past^\prime$
that groups all histories which give rise to the same prediction:
\begin{equation}
\epsilon(\past) =
  \{ \past^\prime: \Pr(\Future|\past) = \Pr(\Future|\past^\prime) \} ~. 
\label{Eq:PredictiveEquivalence}
\end{equation}
In other words, for the purpose of forecasting the future, two different pasts
are equivalent if they result in the same prediction. The result of applying
this equivalence gives the process's \emph{causal states} 
$\CausalStateSet = \Pr(\Past,\Future) / \sim$, which partition the space
$\AllPasts$ of pasts into sets that are predictively equivalent.
The set of causal states
\footnote{A process's causal states consist of both transient and recurrent
states. To simplify the presentation, we henceforth refer \emph{only} to
recurrent causal states that are discrete.}
can be discrete, fractal, or continuous;
see, e.g., Figs. 7, 8, 10, and 17 in \refcite{Crut92c}.

State-to-state transitions are denoted by matrices
$T_{\CausalState \CausalState^\prime}^{(\meassymbol)}$ whose elements give the
probability $\Pr(\MeasSymbol=\meassymbol,\CausalState^\prime|\CausalState)$ of transitioning
from one state $\CausalState$ to the next $\CausalState^\prime$ on seeing
measurement $\meassymbol$. The resulting model, consisting of the causal
states and transitions, is called the process's \emph{\eM}. Given a process
$\mathcal{P}$, we denote its \eM\ by $M(\mathcal{P})$.
 
Causal states have a Markovian property that they render the past and future
statistically independent; they \emph{shield} the future from the past~\cite{Crut98d}:
\begin{equation}
\Pr(\Past,\Future|\CausalState)
  = \Pr(\Past|\CausalState) \Pr(\Future|\CausalState) ~.
\label{shield}
\end{equation}
Moreover, they are optimally predictive~\cite{Crut88a} in the sense that
knowing which causal state a process is in is just as good as having the
entire past: $\Pr(\Future|\CausalState) = \Pr(\Future|\Past)$. In other
words, causal shielding is equivalent to the fact~\cite{Crut98d} that the
causal states capture all of the information shared between past and future:
$I[\CausalState;\Future] = \EE$.

\EMs\ have an important, if subtle, structural property called
\emph{unifilarity}~\cite{Crut88a,Shal98a}: From the start state, each observed 
sequence $\ldots x_{-3} x_{-2} x_{-1} \ldots$ corresponds to one and only one 
sequence of causal states \footnote{Following terminology in computation theory 
this is referred to as \emph{determinism}~\cite{Hopc79}. However, to reduce 
confusion, here we adopt the practice in information theory to call it the 
\emph{unifilarity} of a process's representation~\cite{Ephr02a}.}.
\EM\ unifiliarity underlies many of the results here. Its importance is
reflected
in the fact that representations without unifilarity, such as general hidden
Markov models, \emph{cannot} be used to directly calculate important system
properties---including the most basic, such as, how random a process is.
Nonetheless, unifilarity is easy to verify: For each state, each measurement 
symbol appears on at most one outgoing transition \footnote{Specifically, the 
transition matrices have at most one nonzero component in each row.}.
The signature of unifilarity is that on knowing the current state and 
measurement, the uncertainty in the next state vanishes:
$H[\CausalState_{t+1}|\CausalState_t,\MeasSymbol_t] = 0$.
In summary, a process's \eM\ is its unique minimal unifilar model. 

\section{Information Processing Invariants}

Out of all optimally predictive models $\PrescientStateSet$---for which
$I[\PrescientState;\Future] = \EE$---the \eM\ captures the minimal 
amount of information that a process must store in order to communicate all
of the excess entropy from the past to the future.  This is the Shannon
information contained in the causal states---the \emph{statistical complexity}~\cite{Crut98d}: 
$\Cmu \equiv H[\CausalState] \leq H[\PrescientState]$. In short, $\EE$ is the
effective information transmission rate of the process, viewed 
as a channel, and $\Cmu$ is the sophistication of that channel.

Combined, these properties mean that the \eM\ is the basis against which
modeling should be compared, since it captures all of a process's
information at maximum representational efficiency.

In addition to $\EE$ and $\Cmu$, another key (and historically prior)
invariant for dynamical systems and stochastic processes is the entropy rate:
\begin{equation}
\hmu = \lim_{L \rightarrow \infty} \frac{H(L)}{L} ~,
\end{equation}
where $H(L)$ is Shannon entropy of length-$L$ sequences $\MeasSymbol^L$. This
 is the per-measurement rate at which the process generates
information---its degree
of intrinsic randomness~\cite{Shan62,Kolm58}.

Importantly, due to unifilarity one can calculate the entropy rate directly
from a process's \eM:
\begin{align}
\hmu
  & = H[\MeasSymbol|\CausalState] \nonumber \\
  & = - \sum_{\{\CausalState\}} \Pr(\CausalState)
  \sum_{\{\meassymbol\}} T^{(\meassymbol)}_{\CausalState\CausalState^\prime}
  \log_2 T^{(\meassymbol)}_{\CausalState\CausalState^\prime}
  ~  .
\label{eq:hmuEM}
\end{align}
$\Pr(\CausalState)$ is the asymptotic
probability of
the causal states, which is obtained as the normalized principal eigenvector
of the transition matrix $T = \sum_{\{x\}} T^{(x)}$.  We will use $\pi$ 
to denote the distribution over the causal states as a row vector. Note that 
a process's statistical complexity can also be directly calculated
from its \eM:
\begin{align}
\Cmu
   & = H[\CausalState] \nonumber \\
   & = - \sum_{\{\CausalState\}} \Pr(\CausalState) \log_2 \Pr(\CausalState) ~.
\end{align}
Thus, the \eM\  directly gives two important invariants: a process's rate
($\hmu$) of producing information and the amount ($\Cmu$) of historical
information it stores in doing so.

\section{Excess Entropy}

Until recently, $\EE$ could not be as directly calculated 
as the entropy rate and the statistical complexity. This state of affairs 
was a major roadblock to analyzing the relationships between modeling
and predicting and, more concretely, the relationships between (and even the
interpretation of) a process's basic invariants---$\hmu$, $\Cmu$, and $\EE$.
\refcite{Crut08a} announced the solution to this longstanding problem by 
deriving explicit expressions for $\EE$ in terms of the \eM, providing a 
unified information-theoretic analysis of general processes. Here we
provide a detailed account of the underlying methods and results.

To get started, we should recall what is already known about the relationships
between these various quantities.
First, some time ago, an explicit expression was developed from the Hamiltonian
for one-dimensional spin chains with range-$R$ interactions~\cite{Crut97a}:
\begin{equation}
\EE = \Cmu - R \, \hmu ~.
\label{SpinEEandCmu}
\end{equation}
It was demonstrated that $\EE$ is a generalized order parameter: Compared to
structure factors, $\EE$ is an assumption-free way to find structure and
correlation in spin systems that does not require tuning~\cite{Feld98b}.

Second, it has also been known for some time that the statistical complexity
is an upper bound on the excess entropy~\cite{Shal98a}:
\begin{equation}
\EE \leq \Cmu ~.
\label{CmuEBound}
\end{equation}
Nonetheless, other than the special, if useful, case of spin systems, until
\refcite{Crut08a} there had been no direct way to calculate $\EE$. Remedying
this limitation required broadening the notion of what a process is.

\section{Retrodiction}

The original results of computational mechanics concern using the past to
predict the future. But we can also retrodict: use the future to predict
the past. That is, we scan the measurement variables not in the forward time
direction, but in the reverse. The computational mechanics formalism is
essentially unchanged, though its meaning and notation need to be augmented
\cite{Crut91b}.

With this in mind, the previous mapping from pasts to causal states is
now denoted $\FutureEps$ and it gave, what we will call, the
\emph{predictive} causal states
$\FutureCausalStateSet$. When scanning in the reverse direction, we
have a new relation, $\future \PastSim \future^\prime$, which groups futures
that are equivalent for the purpose of retrodicting the past:
$\PastEps(\future) =
  \{ \future^\prime: \Pr(\Past|\future) = \Pr(\Past|\future^\prime) \}$.
It gives the \emph{retrodictive} causal states
$\PastCausalStateSet = \Pr(\Past,\Future) / \PastSim$.
And, not surprisingly, we must also distinguish the forward-scan
\eM\ $\FutureEM$ from the reverse-scan \eM\ $\PastEM$. They assign
corresponding entropy rates, $\Futurehmu$ and $\Pasthmu$, and
statistical complexities, $\FutureCmu = H[\FutureCausalState]$
and $\PastCmu = H[\PastCausalState]$,
respectively, to the process.

To orient ourselves, a graphical aid, the \emph{hidden process lattice}, is
helpful at this point; see \tableref{tab:ProcessLattice}.

\renewcommand\arraystretch{1.65} 
\begin{table}[th]
\begin{center}
\begin{tabular}{ccccccc|c|ccccccc}
\hline
&& && && Past & Present & \rlap{Future}\phantom{Past}\\
\hline
&& && && $\Past$ &   & $\Future$ \\
$\ldots$ && $ \negthickspace \MeasSymbol_{-3} \negthickspace $ &&
$ \negthickspace \MeasSymbol_{-2} \negthickspace $ &&
	$ \negthickspace \MeasSymbol_{-1} \negthickspace $ & &
	$ \negthickspace \MeasSymbol_0 \negthickspace $ && $
	\negthickspace \MeasSymbol_1 \negthickspace $ && $
	\negthickspace \MeasSymbol_2 \negthickspace $ && $
	\negthickspace \ldots$ \\
\hline
$\ldots$ & \negthickspace $\FutureCausalState_{-3} \negthickspace$
	&& $\negthickspace \FutureCausalState_{-2} \negthickspace$
	&& $\negthickspace \FutureCausalState_{-1} \negthickspace$
	&& $\FutureCausalState_{0}$ &&
	$ \negthickspace \FutureCausalState_{1} \negthickspace $ && $
	\negthickspace \FutureCausalState_{2} \negthickspace $ && $
	\negthickspace \FutureCausalState_{3} \negthickspace $
	& $ \negthickspace \ldots$ \\
$\ldots$ & \negthickspace $\PastCausalState_{-3} \negthickspace$
	&& $\negthickspace \PastCausalState_{-2} \negthickspace$
	&& $\negthickspace \PastCausalState_{-1} \negthickspace$
	&& $\PastCausalState_{0}$ &&
	$ \negthickspace \PastCausalState_{1} \negthickspace $ && $
	\negthickspace \PastCausalState_{2} \negthickspace $ && $
	\negthickspace \PastCausalState_{3} \negthickspace $
	& $ \negthickspace \ldots$ \\
\hline
\end{tabular}
\renewcommand\arraystretch{1.0} 
\end{center}
\caption{
  Hidden Process Lattice: The $\MeasSymbol$ variables denote the
  observed process; the $\CausalState$ variables, the hidden
  states. If one scans the observed variables in the
  positive direction---seeing $\MeasSymbol_{-3}$, $\MeasSymbol_{-2}$,
  and $\MeasSymbol_{-1}$---then that history takes one to causal
  state $\FutureCausalState_0$. Analogously, if one scans in the reverse
  direction, then the succession of variables $\MeasSymbol_{2}$,
  $\MeasSymbol_{1}$, and $\MeasSymbol_{0}$ leads to $\PastCausalState_0$.
  }
\label{tab:ProcessLattice}
\end{table}

Now we are in a position to ask some questions. Perhaps the most obvious is,
In which time direction is a process most predictable? The answer is that a
process is equally predictable in either:
\begin{Prop}
\cite{Crut98d}
For a stationary process, optimally predicting the future
and optimally retrodicting the past are equally effective:
$\Pasthmu = \Futurehmu$.
\label{prop:FuturePastEqPredict}
\end{Prop}

\begin{ProProp}
A stationary stochastic process satisfies:
\begin{equation}
H[X_{-L+2},\ldots,\MeasSymbol_0] = H[X_{-L+1},\ldots,\MeasSymbol_{-1}] ~.
\end{equation}
Keeping this in mind, we directly calculate:
\begin{align*}
\Futurehmu & = H[\MeasSymbol_0|\Past] \\
  & = \lim_{L \rightarrow \infty}
  	H[\MeasSymbol_0|\MeasSymbol_{-L+1},\ldots,\MeasSymbol_{-1}] \\
  & = \lim_{L \rightarrow \infty}
    \left( H[\MeasSymbol_{-L+1},\ldots,\MeasSymbol_0]
		- H[\MeasSymbol_{-L+1},\ldots,\MeasSymbol_{-1}] \right) \\
  & = \lim_{L \rightarrow \infty}
    \left( H[\MeasSymbol_{-L+1},\ldots,\MeasSymbol_0]
		- H[\MeasSymbol_{-L+2},\ldots,\MeasSymbol_0] \right) \\
  & = \lim_{L \rightarrow \infty}
    \left( H[\MeasSymbol_{-1},\ldots,\MeasSymbol_{L-2}]
		- H[\MeasSymbol_0,\ldots,\MeasSymbol_{L-2}] \right) \\
  & = \lim_{L \rightarrow \infty}
  	H[\MeasSymbol_{-1} | \MeasSymbol_0,\ldots,\MeasSymbol_{L-2}] \\
  & = H[\MeasSymbol_{-1}|\Future] \\
  & = \Pasthmu ~. \qed 
\end{align*}
\end{ProProp}

Somewhat surprisingly, the effort involved in optimally predicting and
retrodicting is not necessarily the same:
\begin{Prop}
\cite{Crut91b}
There exist stationary processes for which $\PastCmu \neq \FutureCmu$.
\label{ProcessNotTimeSymmetric}
\end{Prop}

\begin{ProProp}
The random-insertion process, analyzed in a later section,
establishes this by example.
\end{ProProp}

Note that $\EE$ is mute on this score. Since the mutual information $I$ is
symmetric in its variables~\cite{Cove06a}, $\EE$ is time symmetric.
Proposition \ref{ProcessNotTimeSymmetric} puts us on notice that $\EE$
necessarily misses many of a process's structural properties.

\section{Excess Entropy from Causal States}

The relationship between predicting and retrodicting a process, and ultimately
$\EE$'s role, requires teasing out how the states of the forward and reverse
\eMs\ capture information from the past and the future. To do this we
analyzed~\cite{Crut08c} a four-variable mutual information:
$I[\Past;\Future;\FutureCausalState;\PastCausalState]$.
A large number of expansions of this quantity are possible. A systematic
development follows from \refcite{Yeun91a} which showed that Shannon entropy
$H[\cdot]$ and mutual information $I[\cdot;\cdot]$ form a signed measure over
the space of events. Practically, there is a direct correspondence between set
theory and these information measures. Using this, \refcite{Crut08c} developed
an \emph{\eM\ information diagram} over four variables, which gives a minimal
set of entropies, conditional entropies, mutual informations, and conditional
mutual informations necessary to analyze the relationships among $\hmu$,
$\Cmu$, and $\EE$ for general stochastic processes.

In a generic four-variable information diagram, there are 15 independent
variables. Fortunately, this greatly simplifies in the case of using an \eM\ to
represent a process; there are only 5 independent variables in the \eM\
information diagram~ \cite{Crut08c}. (These results are announced
in \cite{Crut08a}; see Fig. 1 there.)

Simplified in this way, we are left with our main results which, due to the
preceding effort, are particularly transparent.
\begin{The}
Excess entropy is the mutual information between the predictive and
retrodictive causal states:
\begin{equation}
\EE = I[\FutureCausalState;\PastCausalState] ~.
\end{equation}
\label{EasCausalMI}
\end{The}

\begin{ProThe}
This follows due to the redundancy of pasts and predictive causal states, on
the one hand, and of futures and retrodictive causal states, on the other.
These redundancies, in turn, are expressed via
$\FutureCausalState = \FutureEps(\Past)$ and
$\PastCausalState = \PastEps(\Future)$, respectively. That is, we have
\begin{align}
I[\Past;\Future;\FutureCausalState;\PastCausalState]
	& = I[\Past;\Future] \nonumber \\
	& = \EE ~,
\end{align}
on the one hand, and
\begin{equation}
I[\Past;\Future;\FutureCausalState;\PastCausalState]
	= I[\FutureCausalState;\PastCausalState] ~,
\end{equation}
on the other.
\qed
\end{ProThe}

That is, the process's effective channel capacity \mbox{$\EE = I[\Past;\Future]$}
is the same as that of a ``channel'' between the forward and reverse
\eM\ states.

\begin{Prop}
The predictive and retrodictive statistical complexities are:
\begin{align}
\label{FutureCmuReln}
\FutureCmu &= \EE + H[\FutureCausalState|\PastCausalState]
  \mathrm{~and} \\
\label{PastCmuReln}
\PastCmu &= \EE + H[\PastCausalState|\FutureCausalState] ~.
\end{align}
\end{Prop}
    
\begin{ProProp}
$\EE = I[\FutureCausalState;\PastCausalState] =
H[\FutureCausalState] - H[\FutureCausalState|\PastCausalState]$.
Since the first term is $\FutureCmu$, we have the predictive
statistical complexity. Similarly for the retrodictive complexity.
\qed
\end{ProProp}
    
\begin{Cor}
$\FutureCmu \geq H[\FutureCausalState|\PastCausalState]$ and 
$\PastCmu \geq H[\PastCausalState|\FutureCausalState]$.
\end{Cor}
    
\begin{ProCor}
$\EE \geq 0$.
\end{ProCor}
    
The Theorem and its companion Proposition give an explicit connection between
a process's excess entropy and its causal structure---its \eMs. More generally,
the relationships directly tie mutual information measures of observed
sequences to a process's internal structure. This is our main result. It allows
us to probe the properties that control how closely observed statistics reflect
a process's hidden organization. However, this requires that we understand how 
$\FutureEM$ and $\PastEM$ are related. We express this relationship with a unifying
model---the bidirectional machine.

\section{The Bidirectional Machine}

At this point, we have two separate \eMs---one for predicting ($\FutureEM$) and
one for retrodicting ($\PastEM$). We will now show that one can do better, 
by simultaneously utilizing causal information from the past and future.

\begin{Def}
Let $\BiEM$ denote the \emph{bidirectional machine} given by the equivalence
relation $\BiEquiv$ \footnote{Interpret the symbol $\pm$
as ``plus \emph{and} minus''.}:
\begin{align*}
\BiEps(\biinfinity) & = \BiEps(\past,\future) \\
  & = \{ (\pastprime,\futureprime):
  	\pastprime \in \FutureEps(\past) ~\mathrm{and}~
	\futureprime \in \PastEps(\future) \}
\end{align*}
with causal states $\BiCausalStateSet = \Pr(\BiInfinity) / \BiEquiv$.
\end{Def}

That is, the bidirectional causal states are a partition of $\BiInfinity$:
$\BiCausalStateSet \subseteq \FutureCausalStateSet \times \PastCausalStateSet$.
This follows from a straightforward adaptation of the analogous result for
forward \eMs~\cite{Crut98d}.

To illustrate, imagine being given a particular realization $\biinfinity$. In
effect, the bidirectional machine $\BiEM$ describes how one can move around on
the hidden process lattice of \tableref{tab:ProcessLattice}:
\begin{enumerate}
\item When scanning in the forward direction, states and transitions associated
	with $\FutureEM$ are followed.
\item When scanning in the reverse direction, states and transitions associated
	with $\PastEM$ are followed.
\item At any time, one can change to the opposite scan direction, moving to
	the state of the opposite scan's \eM. For example, if one moves forward
	following $\FutureEM$
	and ends in state $\FutureCausalState$, having seen $\past$ and about to
	see $\future$, then one moves to $\PastCausalState = \PastEps(\future)$.
\end{enumerate}
At time $t$, the bidirectional causal state is
$\BiCausalState_t = (\FutureEps(\past_t),\PastEps(\future_t))$. When scanning
in the forward direction, the first symbol of $\future_t$ is removed and
appended to $\past_t$. When scanning in the reverse direction,
the last symbol in $\past_t$ is removed and prefixed to $\future_t$. In either
situation, the new bidirectional causal state is determined by $\BiEps$ and
the updated past and future.

This illustrates the relationship between $\FutureCausalState$ and
$\PastCausalState$, as specified by $\BiEM$, when given a particular
realization. Generally, though, one considers an ensemble $\BiInfinity$ of
realizations. In this case, the bidirectional state transitions are
probabilistic and possibly nonunifilar. This relationship can be made more
explicit through the use of maps between the forward and reverse causal states.
These are the \emph{switching} maps.

The forward map is a linear function from the simplex over $\PastCausalStateSet$
to the simplex over $\FutureCausalStateSet$, and analogously for the reverse
map. The maps are defined in terms of conditional probability distributions:
\begin{enumerate}
\item The \emph{forward map} $\ForwardMap: \Delta^n \rightarrow \Delta^m$, where
$\ForwardMap(\pastcausalstate) = \Prob(\FutureCausalState | \pastcausalstate)$;
and
\item The \emph{reverse map} $\ReverseMap: \Delta^m \rightarrow \Delta^n$,
where
$\ReverseMap(\futurecausalstate) = \Prob(\PastCausalState | \futurecausalstate)$,
\end{enumerate}
where $n = |\PastCausalStateSet|$ and $m = |\FutureCausalStateSet|$.

We will sometimes refer to these maps in the Boolean rather than probabilistic
sense. The case will be clear from context.

\begin{Prop}
$\ReverseMap$ and $\ForwardMap$ are onto.
\label{SwitchingMapsAreOnto}
\end{Prop}

\begin{ProProp}
Consider the reverse map $\ReverseMap$ that takes one from a forward causal
state to a reverse causal state. Assume $\ReverseMap$ is not onto. 
Then there must be a reverse state $\pastcausalstate$
that is not in the range of $r(\FutureCausalState)$. This means that no forward
causal state is paired with $\pastcausalstate$ and so there is no past $\past$
with a possible future $\future \in \pastcausalstate$. That is,
$\BiEps(\past,\future) = \emptyset$ and, specifically,
$\PastEps(\future) = \emptyset$.
Thus, $\pastcausalstate$ does not exist.

A similar argument shows that $\ForwardMap$ is onto.
\qed
\end{ProProp}

\begin{Def}
The amount of stored information needed to optimally predict and retrodict
a process is $\BiEM$'s statistical complexity:
\begin{equation}
\BiCmu \equiv H[\BiCausalState] = H[\FutureCausalState,\PastCausalState] ~.
\end{equation}
\end{Def}

From the immediately preceding results we obtain the following simple,
explicit, and useful relationship:
\begin{Cor}
$\EE = \FutureCmu + \PastCmu - \BiCmu$.
\end{Cor}

Thus, we are led to a wholly new interpretation of the excess entropy---in
addition to the original three discussed in \refcite{Crut01a}: $\EE$ is
exactly the difference between these structural complexities. Moreover,
only when \mbox{$\EE = 0$} does $\BiCmu = \FutureCmu + \PastCmu$.

More to the point, thinking of the $\Cmu$s as proportional to the size of the
corresponding machine, we establish the representational efficiency of the
bidirectional machine:
\begin{Prop}
$\BiCmu \leq \FutureCmu + \PastCmu$.
\end{Prop}

\begin{ProProp}
This follows directly from the preceding corollary and the nonnegativity of
mutual information.
\qed
\end{ProProp}

We can say a bit more, with the following bounds.

\begin{Cor}
$\FutureCmu \leq \BiCmu$ and $\PastCmu \leq \BiCmu$.
\end{Cor}

These results say that taking into account causal information from the
past \emph{and} the future is more efficient (i) than ignoring one or the
other and (ii) than ignoring their relationship.

\subsection{Upper Bounds}

Here we give new, tighter bounds for $\EE$ than Eq. (\ref{CmuEBound}) and
greatly simplified proofs than those provided in Refs.~\cite{Crut98d} and
\cite{Shal98a}.

\begin{Prop}
For a stationary process, $\EE \leq \FutureCmu$ and $\EE \leq \PastCmu$.
\end{Prop}

\begin{ProProp}
These bounds follow directly from applying basic information inequalities:
$I[X,Y] \leq H[X]$ and $I[X,Y] \leq H[Y]$.
Thus, $\EE = I[\PastCausalState;\FutureCausalState] \leq
H[\PastCausalState]$, which is $\PastCmu$. Similarly, since
$I[\PastCausalState;\FutureCausalState] \leq 
H[\FutureCausalState]$, we have $\EE \leq \FutureCmu$.
\qed
\end{ProProp}

\subsection{Causal Irreversibility}

We have shown that predicting and retrodicting may require different amounts of
information storage (\mbox{$\FutureCmu \neq \PastCmu$}). We now examine this asymmetry.

Given a word $w = x_0 x_2 \ldots x_{L-1}$, the word we see when scanning in
the reverse direction is $\widetilde{w} = x_{L-1} \ldots x_1 x_0$, where
$x_{L-1}$ is encountered first and $x_0$ is encountered last.

\begin{Def}
A \emph{microscopically reversible process} is one for which
$\Pr(w) = \Pr(\widetilde{w})$, for all words $w = x^L$ and all $L$.
\end{Def}

Microscopic reversibility simply means that flipping $t \rightarrow -t$ leads
to the same process $\Pr(\Past,\Future)$. A microscopically
reversible process scanned in both directions yields the same word
distribution; we will denote this $\FutureProcess = \PastProcess$.

\begin{Prop}
A microscopically reversible process has $\PastEM = \FutureEM$.
\end{Prop}

\begin{ProProp}
If $\FutureProcess = \PastProcess$, then $\eMachine(\FutureProcess) = 
\eMachine(\PastProcess)$ since \eMachine is a function. And these
are $\FutureEM$ and $\PastEM$, respectively.
\qed
\end{ProProp}

\begin{Cor}
For a microscopically reversible process, $\PastCmu = \FutureCmu$.
\end{Cor}

\begin{ProCor}
For a microscopically reversible process $\PastEM = \FutureEM$. And so, in
particular, $\PastCausalStateSet = \FutureCausalStateSet$, their transition
matrices are the same, and so
$\Pr(\PastCausalState) = \Pr(\FutureCausalState)$.
Thus, $\PastCmu = \FutureCmu$.
\qed
\end{ProCor}

Now consider a slightly looser, and more helpful, notion of reversibility,
expressed quantitatively as a measure of irreversibility.

\begin{Def}
A process's \emph{causal irreversibility}~\cite{Crut91b} is:
\begin{equation}
\CI(\Process) = \FutureCmu - \PastCmu ~.
\end{equation}
\end{Def}

\begin{Cor}
$\CI(\Process) = H[\FutureCausalState|\PastCausalState] -
H[\PastCausalState|\FutureCausalState]$.
\end{Cor}

Note that $\CI = 0$ does not imply that $\FutureEM = \PastEM$.
For example, the periodic process $\ldots 123123123 \ldots$ is not
microscopically reversible, since $\Pr(123) \neq \Pr(321)$.
However, $\CI = 0$, as $\PastCmu = \FutureCmu = \log_2 3$.

It turns out, though, that we are more interested in the following situation.

\begin{Prop}
If $\CI(\Process) \neq 0$, then the process is not microscopically reversible.
\label{CI_MicroIrreversible}
\end{Prop}

\begin{ProProp}
$\FutureCmu \neq \PastCmu$ implies that $\FutureEM \neq \PastEM$.
And so, $\FutureProcess \neq \PastProcess$.
\qed
\end{ProProp}

So, a vanishing $\CI$ will indicate ``reversibility'' for some classes of
processes that are not microscopically reversible.  The periodic process just 
described is one such example. In fact, this includes any process whose 
left- and right-scan processes are isomorphic under a simultaneous 
measurement-alphabet and causal-state isomorphism. Given that the 
spirit of symbolic dynamics is to consider processes only up to 
isomorphism, this measure seems to capture a very natural notion of
irreversibility. Interestingly, it appears, based on several case studies, 
that causal reversibility captures \emph{exactly} that notion. That is, it would 
seem there are no processes for which $\CI=0$, yet $\FutureProcess \nsim 
\PastProcess$. We leave this as a conjecture.


Finally, note that causal irreversibility is not controlled by $\EE$, since,
as noted above, the latter is scan-symmetric.

\subsection{Process Crypticity}

Lurking in the preceding development and results is an alternative
view of how forecasting and modeling building are related.

We can extend our use of Shannon's communication theory (processes are memoryful
channels) to view the activity of an observer building a model of a process as
the attempt to decrypt from a measurement sequence the hidden state information
\cite{Shan49a}. The parallel we draw is that the design goal of cryptography
is to not reveal internal correlations and structure within an encrypted data
stream, even though in fact there is a message---hidden organization and
structure---that will be revealed to a recipient with the correct codebook.
This is essentially the circumstance a scientist faces when building a model,
for the first time, from measurements: What are the states and dynamic
(hidden message) in the observed data?

Here, we address only the case of \emph{self-decoding} in which the information
used to build a model is only that available in the observed process
$\Pr(\BiInfinity)$. That is, no ``side-band'' communication, prior knowledge,
or disciplinary assumptions are allowed. Note, though, that modeling with such
additional knowledge requires solving the self-decoding case, addressed here,
first. The self-decoding approach to building nonlinear models from time series
was introduced in \refcite{Pack80}.

The relationship between excess entropy and statistical complexity established
by Thm. \ref{EasCausalMI} indicates that there are fundamental limitations on
the amount of a process's stored information directly present in
observations, as reflected in the mutual information measure $\EE$.
We now introduce a measure of this accessibility.

\begin{Def} A process's \emph{crypticity} is:
\begin{equation}
\PC(\FutureEM,\PastEM) =
	H[\FutureCausalState|\PastCausalState]
	+ H[\PastCausalState|\FutureCausalState] ~.
\end{equation}
\end{Def}

\begin{Prop}
$\PC(\FutureEM,\PastEM)$ is the distance between a process's forward and
reverse \eMs.
\label{Prop:PCisaDistance}
\end{Prop}

\begin{ProProp}
$\PC(\FutureEM,\PastEM)$ is nonnegative, symmetric, and satisfies a triangle
inequality. These follow from the solution of exercise 2.9 of \refcite{Cove06a}.
See also, \refcite{Crut87f}.
\qed
\end{ProProp}

\begin{The}
$\BiEM$'s statistical complexity is:
\begin{equation}
\BiCmu = \EE + \PC ~.
\end{equation}
\end{The}

\begin{ProThe}
This follows directly from the corollary and the predictive and retrodictive
statistical complexity relations, Prop. (\ref{FutureCmuReln}) and (\ref{PastCmuReln}).
\qed
\end{ProThe}

Referring to $\PC$ as crypticity comes directly from this result: It
is the amount of internal state information ($\BiCmu$) not locally
present in the observed sequence ($\EE$). That is, a process hides
$\PC$ bits of information.

Note that if crypticity is low $\PC \approx 0$, then much of the stored
information is present in observed behavior: $\EE \approx \BiCmu$. However,
when a process's crypticity is high, $\PC \approx \BiCmu$, then little of it's
structural information is directly present in observations. The measurements
appear very close to being independent, identically distributed
($\EE \approx 0$) despite the fact that the process can be highly structured
($\BiCmu \gg 0$).

\begin{Cor}
$\BiEM$'s statistical complexity bounds the process's crypticity:
\begin{equation}
\BiCmu \geq \PC ~.
\end{equation}
\end{Cor}

\begin{ProCor}
$\EE \geq 0$.
\qed
\end{ProCor}

Thus, a truly cryptic process has $\BiCmu = \PC$ or, equivalently, $\EE = 0$.
In this circumstance, little or nothing can be learned about the process's
hidden organization from measurements. This would be perfect encryption.

We will find it useful to discuss the two contributions to $\PC$ separately.
Denote these \mbox{$\FuturePC = H[\FutureCausalState|\PastCausalState]$} and
\mbox{$\PastPC = H[\PastCausalState|\FutureCausalState]$}.

The preceding results can be compactly summarized in an information diagram
that uses the \eM\ representation of a process; see \refcite{Crut08a} and
\refcite{Crut08c}. They also lead to a new classification scheme for stationary
processes; see \refcite{Crut08d}. In the following, we concentrate instead on
how to calculate the preceding quantities, giving a complete informational and
structural analysis of general processes.

\section{Alternative Presentations}

The \eM\ is a process's unique, minimal unifilar presentation. Now we introduce
two alternative presentations, which need not be \eMs, that will be used in
the calculation of \EE.  Since the states of these alternative presentations
are not causal states, we will use $\AlternateState_t$, rather than 
$\CausalState_t$, to denote the random variable for their state at time $t$.

\subsection{Time-Reversed Presentation}

Any machine $M$ transitions from the current state $\AlternateState$ to the
next state $\AlternateStatePrime$ on the current symbol $\meassymbol$:
\begin{equation}
T^{(x)}_{\AlternateState\AlternateStatePrime} 
  \equiv 
  \Pr(\MeasSymbol=\meassymbol, \AlternateStatePrime | \AlternateState) ~.
\end{equation}
Note that $T = \sum_{\{x\}} T^{(x)}$ is a stochastic matrix with principal
eigenvalue $1$ and left eigenvector $\pi$, which gives $\Pr(\AlternateState)$.
Recall that the Perron-Frobenius theorem applied to stochastic matrices
guarantees the uniqueness of $\pi$.

Using standard probability rules to interchange $\AlternateState$ and 
$\AlternateStatePrime$, we can construct a new set of transition matrices
which defines a presentation of the process that generates the symbols in 
reverse order. It is useful to consider a time-reversing operator acting on a
machine. Denoting it $\TR$, $\widetilde{M} = \TR(M)$ is the
\emph{time-reversed presentation} of $M$. It has symbol-labeled
transition matrices:
\begin{align}
\widetilde{T}^{(x)}_{\AlternateStatePrime\AlternateState}
  & \equiv
\Pr(\MeasSymbol=\meassymbol, \AlternateState | \AlternateStatePrime)
\nonumber \\
  & = 
T^{(x)}_{\AlternateState\AlternateStatePrime} 
    \frac{ \Pr(\AlternateState) }{ \Pr(\AlternateStatePrime) } ~.
\end{align}
and stochastic matrix $\widetilde{T} = \sum_{\{x\}} \widetilde{T}^{(x)}$.

\begin{Prop}
\label{eqpi}
The stationary distribution $\widetilde{\pi}$ over the time-reversed
presentation states is the same as the stationary distribution $\pi$ of $M$.
\end{Prop}

\begin{ProProp}
We assume $\widetilde{\pi} = \pi$, the left eigenvector of $T$, and verify the
assumption, recalling the uniqueness of $\pi$. We have:
\begin{align*}
\widetilde{\pi}_\alternatestate 
  &= 
\sum_{\alternatestateprime} \widetilde{\pi}_{\alternatestate^{\prime}} 
                            \widetilde{T}_{\alternatestate^{\prime}\alternatestate} \\
  &=
\sum_{\alternatestateprime} \widetilde{\pi}_{\alternatestate^{\prime}} 
                            T_{\alternatestate\alternatestate^{\prime}} 
                            \frac{\pi_\alternatestate}{\pi_{\alternatestateprime}} \\
 &= 
\sum_{\alternatestateprime} T_{\alternatestate\alternatestate^{\prime}} 
                            \pi_\alternatestate \\
 &= \pi_\alternatestate ~.
\qed
\end{align*}
In the second to last line, we recall the assumption
$\widetilde{\pi}_\alternatestateprime = \pi_\alternatestateprime$.
And in the final, we note that $T$ is stochastic.
\qed
\end{ProProp}

Finally, when we consider the product
of transition matrices over a given sequence $w$, it is useful to simplify
notation as follows:
\begin{align*}
T^{(w)} \equiv T^{(\meassymbol_0)} T^{(\meassymbol_1)} \cdots 
               T^{(\meassymbol_{L-1})}.
\end{align*}

\subsection{Mixed-State Presentation}
\label{MSP}

The states of machine $M$ can be treated as a standard basis in a vector space.
Then, any distribution over these states is a linear combination of those
basis vectors. Following \refcite{Uppe97a}, these distributions are 
called \emph{mixed states}. 

Now we focus on a special subset of mixed states and define $\mu(w)$ as 
the distribution over the states of $M$ that is induced after observing $w$:
\begin{align}
\label{mixedstates}
\mu(w) 
  &\equiv \Pr(\AlternateState_L | \MeasSymbol_0^L=w)  \\
  & = \frac{\Pr(\MeasSymbol_0^L=w, \AlternateState_L)}{\Pr(\MeasSymbol_0^L=w)} 
\\
  &= \frac{\pi T^{(w)}}{\pi T^{(w)} \one},
\end{align}
where $\MeasSymbol_0^L$ is shorthand for an undetermined sequence of $L$ 
measurements beginning at time $t=0$ and $\one$ is a column vector of $1$s.
In the last line, we write the
probabilities in terms of the stationary distribution and the transition
matrices of $M$.  This expansion is valid for any machine that generates
the process in the forward-scan (left-to-right) direction.

If we consider the entire set of such mixed states, then we can construct a 
presentation of the process by specifying the transition matrices:
\begin{align}
\Pr(\meassymbol, \mu(w\meassymbol) | \mu(w))
  & \equiv \frac{\Pr(wx)}{\Pr(w)} \\
  & = \mu(w) T^{(x)} \one ~.
\end{align}
Note that many words can induce the same mixed state. As with the time-reversed 
presentation, it will be useful to define a
corresponding operator $\MSP$ that acts on a machine $M$, returning its
\emph{mixed-state presentation} $\MSP(M)$.

\section{Calculating Excess Entropy}

We are now ready to describe how to calculate the excess entropy, using the
time-symmetric perspective. Generally, our goal is to obtain a conditional
distribution $\Pr(\FutureCausalState|\PastCausalState)$ which, when combined
with the \eMs, yields a direct calculation of $\EE$ via Thm.~\ref{EasCausalMI}.
This is a two-step procedure which begins with $\FutureEM$, calculates
$\widetilde{M}^+$, and ends with $\PastEM$. One could also start with $\PastEM$
to obtain $\FutureEM$. These possibilities are captured in the diagram:
\begin{equation}
\begin{CD}
\FutureEM @<\MSP<< \widetilde{M}^-\\
@V\TR{}VV             @AA\TR{}A\\
\widetilde{M}^+ @>>\MSP> \PastEM
\end{CD}
\end{equation}
In detail, we begin with $\FutureEM$ and reverse the direction of time by 
constructing the time-reversed presentation $\widetilde{M}^+ = \TR(\FutureEM)$.
Then, we construct the mixed-state presentation $\MSP(\widetilde{M}^+)$ of the
time-reversed presentation to obtain $\PastEM$.

Note that $\TR$ acting on $\FutureEM$ does not generically yield another \eM.
(This was not the purpose of $\TR$.) However, the states will still be useful
when we construct the mixed-state presentation of $\widetilde{M}^+$. This is
because the states, which serve as basis states in the mixed-state presentation,
are in a one-to-one correspondence with the forward causal states of
$\FutureEM$. This correspondence was established by Prop.~\ref{eqpi}.

Also, note that $\MSP$ is not guaranteed to construct a minimal presentation 
of the process. However, this does not appear to be an issue when working with
time-reversed presentations of an \eM. We leave it as a conjecture that
$\MSP(\TR(M))$ is always minimal. Even so, the Appendix demonstrates that an
appropriate sum can be carried out which always yields the desired conditional
distribution.

Returning to the two-step procedure, one must construct the mixed-state 
presentation of $\widetilde{M}^+$. It is helpful to keep the hidden process 
lattice of \tableref{tab:ProcessLattice} in mind.  Since $\widetilde{M}^+$ 
generates the process from right-to-left, it encounters symbols of $w$ in 
reverse order. The consequence of this is that the form of the mixed state 
changes slightly. However, it \textit{still} represents the distribution over
the current state induced by seeing $w$. We denote
this new form by $\nu(w)$:
\begin{align}
\label{mixedstatesrev}
\nu(w) &\equiv
    \Pr(\AlternateState_0 | \MeasSymbol_0^L=w) \\
  & = 
\frac{\Pr(\AlternateState_0, \MeasSymbol_0^L=w)}{\Pr(\MeasSymbol_0^L=w)} \\
  &= \frac{\pi T^{(\widetilde{w})}}
          {\pi T^{(\widetilde{w})}\one} ~,
\end{align}
where $\pi$ and $T$ are the stationary distribution and transition matrices of
a machine that generates the process from right-to-left, respectively. In this
procedure, we are making use of $\widetilde{M}^+$ and thus, $\widetilde{\pi}$
and $\widetilde{T}$.

Similarly, if we consider the entire set of such mixed states, we can construct 
a presentation of the process by specifying the transition matrices:
\begin{align}
\Pr(\meassymbol, \nu(\meassymbol w) | \nu(w))
	& \equiv \frac{\Pr(xw)}{\Pr(w)} \\
  & = \nu(w) T^{(x)} \one.
\label{transitionrev}
\end{align}

Focusing again on $\FutureEM$, we construct $\widetilde{M}^+ = \TR(\FutureEM)$.
Since $\widetilde{\pi} = \pi$, we can equate 
  $\AlternateState_t = \FutureCausalState_t$
and the mixed states $\nu(w)$ are actually
informing us about the causal states in $\FutureEM$:
\begin{align*}
\nu(w) &= \Pr(\AlternateState_0 | \MeasSymbol_0^L=w)\\
       &= \Pr(\FutureCausalState_0 | \MeasSymbol_0^L=w) ~.
\end{align*}
Whenever the mixed-state presentation is an \eM, each distribution corresponds
to exactly one reverse causal state.  Thus, if $w$ induces $\nu(w)$, then
$\nu(w)$ is the reverse causal state induced by $w$.  This allows us to 
reduce the form of $\nu(w)$ even further so that the conditioned variable 
is a reverse causal state. Continuing,
\begin{align*}
\nu(w) &= \Pr(\FutureCausalState_0 | \MeasSymbol_0^L=w)\\
       &= \Pr\left(\FutureCausalState_0 | \PastCausalState_0 = \PastEps(w)\right).
\end{align*}
Hence, we can calculate $H[\FutureCausalState|\PastCausalState]$ and 
so obtain $\EE$.

\section{Calculational Example}

To clarify the procedure, we apply it to the Random, Noisy Copy (RnC) Process.
The emphasis is on the various process presentations and mixed states that
are used to calculate the excess entropy. In the next section, additional
examples are provided which skip over these calculational details and, instead,
focus on the analysis and interpretation.

The RnC generates a random bit with bias $p$. If that bit is a $0$, it is
copied so that the next output is also $0$. However, if the bit is a $1$, 
then with probability $q$, the $1$ is not copied and $0$ is output 
instead. The RnC Process is related to the \emph{binary asymmetric channel}
of communication theory~\cite{Cove06a}.

\begin{figure}[th]
\centering
\includegraphics[scale=\figscale]{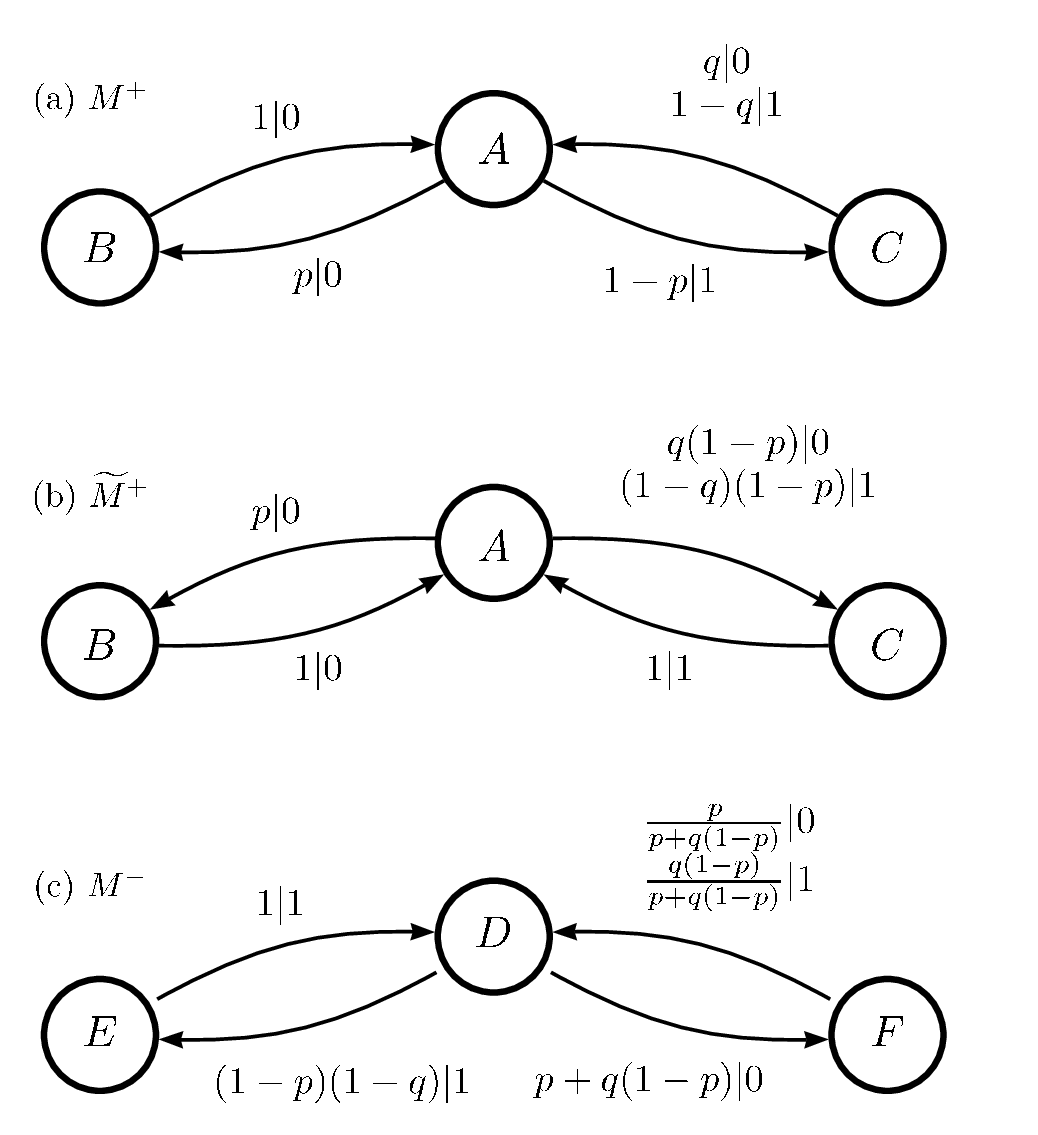}
\caption{
  The presentations used to calculate the excess entropy for the RnC
  Process: (a)~$\FutureEM$, (b)~%
           \mbox{$\widetilde{M}^+ = \TR(\FutureEM)$}, and (c)~%
           \mbox{$\PastEM = \MSP(\widetilde{M}^+)$}.
  Edge labels $t|x$ give the probability
  \mbox{$t = T_{\AlternateState \AlternateState^\prime}^{(x)}$} of making a 
  transition and seeing symbol $x$.
  }
\label{fig:RnC}
\end{figure}

The forward \eM\ has three recurrent causal states 
$\FutureCausalStateSet = \{A,B,C\}$ and is shown in \figref{fig:RnC}(a).
The transition matrices $T^{(x)}$ specify
$\Pr(\MeasSymbol_0=x, \FutureCausalState_1|\FutureCausalState_0)$
and are given by:
\begin{equation*}
T^{(0)} = 
\bordermatrix{ 
  & A & B & C\cr
A & 0 & p & 0\cr
B & 1 & 0 & 0\cr
C & q & 0 & 0
} \\
\end{equation*}
and
\begin{equation*}
T^{(1)} = 
\bordermatrix{ 
  & A & B & C\cr
A & 0 & 0 & 1-p\cr
B & 0 & 0 & 0\cr
C & 1-q & 0 & 0
} ~.
\end{equation*}
(One must explicitly calculate the equivalence classes of histories $\{\past\}$
specified in Eq. (\ref{Eq:PredictiveEquivalence}) and their associated future
conditional distributions $\Pr(\Future|\past)$ to obtain the \eM\ causal states
and transitions.)

These matrices are used calculate the stationary distribution $\pi$ over the 
causal states, which is given by the left eigenvector of the stochastic
matrix $T \equiv T^{(0)} + T^{(1)}$:
\begin{align*}
\Pr(\FutureCausalState) = \frac{1}{2} \bordermatrix{
  & A & B & C\cr
  & 1 & p & 1-p
} ~.
\end{align*}
Using the $T^{(x)}$ and $\pi$, we create the time-reversed presentation
$\widetilde{M}^+ = \TR(\FutureEM)$. This is shown in \figref{fig:RnC}(b).
Notice that the machine is not unifilar, and so it is clearly not an \eM.
The transition matrices for the time-reversed presentation are given by:
\begin{align*}
\widetilde{T}^{(0)} &=
\bordermatrix{ 
  & A & B & C\cr
A & 0 & p & q(1-p)\cr
B & 1 & 0 & 0\cr
C & 0 & 0 & 0
} \mathrm{~and}\\
\widetilde{T}^{(1)} &=
\bordermatrix{ 
  & A & B & C\cr
A & 0 & 0 & (1-q)(1-p)\cr
B & 0 & 0 & 0\cr
C & 1 & 0 & 0
}~.
\end{align*}
As with $M^+$, we calculate the stationary distribution of $\widetilde{M}^+$,
denoted $\widetilde{\pi}$. However, we showed that the stationary
distributions for $M$ and $\TR(M)$ are identical.

Now we are in a position to calculate the mixed-state presentation, 
$\PastEM = \MSP(\widetilde{M}^+)$, shown  in \figref{fig:RnC}(c).  Generally,
causal states can be categorized into types~\cite{Uppe97a}. Of these, 
the calculation of $\EE$ depends only on the reachable recurrent causal states.
The construction of the mixed-state presentation will generate other types of
causal states, such as transient causal states, but we eventually remove
them.

To begin, we start with the empty word, $w=\lambda$, and append $0$ and $1$
to consider $\nu(0)$ and $\nu(1)$, respectively, and calculate:
\begin{align*}
\nu(0) 
  &= \Pr(\FutureCausalState_0|\MeasSymbol_0=0)  \\
  & = \frac{\widetilde{\pi} \widetilde{T}^{(0)}}
 	{\widetilde{\pi} \widetilde{T}^{(0)} \one} \\
  &= \frac{\bigl(p, p, q(1-p)\bigr)}{2p + q(1-p)}
\end{align*}
and
\begin{align*}
\nu(1)
  &= \Pr(\FutureCausalState_0|\MeasSymbol_0=1)  \\
  &= \frac{\widetilde{\pi} \widetilde{T}^{(1)}}
  	{\widetilde{\pi} \widetilde{T}^{(1)} \one} \\
  &= \frac{\bigl( 1, 0, 1-q\bigr)}{2-q} ~.
\end{align*}
For each mixed state, we append $0$s and $1$s and calculate again:
\begin{align*}
\nu(00)
&= \Pr(\FutureCausalState_0|\MeasSymbol_0^2=00) 
 = \frac{\widetilde{\pi} \widetilde{T}^{(0)} \widetilde{T}^{(0)}}
        {\widetilde{\pi} \widetilde{T}^{(0)} \widetilde{T}^{(0)} \one} ~,\\
\nu(01)
&= \Pr(\FutureCausalState_0|\MeasSymbol_0^2=01) 
 = \frac{\widetilde{\pi} \widetilde{T}^{(1)} \widetilde{T}^{(0)}}
        {\widetilde{\pi} \widetilde{T}^{(1)} \widetilde{T}^{(0)} \one} ~,\\
\nu(10)
&= \Pr(\FutureCausalState_0|\MeasSymbol_0^2=10) 
 = \frac{\widetilde{\pi} \widetilde{T}^{(0)} \widetilde{T}^{(1)}}
        {\widetilde{\pi} \widetilde{T}^{(0)} \widetilde{T}^{(1)} \one} ~,
		\mathrm{~and} \\
\nu(11)
&= \Pr(\FutureCausalState_0|\MeasSymbol_0^2=11) 
 = \frac{\widetilde{\pi} \widetilde{T}^{(1)} \widetilde{T}^{(1)}}
        {\widetilde{\pi} \widetilde{T}^{(1)} \widetilde{T}^{(1)} \one }~.
\end{align*}
Note that 
\begin{equation}
\nu(10)
  = \frac{\nu(0)\widetilde{T}^{(1)}}{\nu(0) \widetilde{T}^{(1)}\one} ~.
\end{equation}
This latter form is important in that it allows us to build mixed states from
prior mixed states by prepending a symbol. 

One continues constructing mixed states of longer and longer words until no
more new mixed states appear. As an example, $\nu(1001) = \nu(111001)$ for 
the right-scanned RnC Process.

To illustrate calculating the transition probabilities, consider the
transition from $\nu(00)$ to $\nu(100)$~\footnote{This calculation gives
the probability of transitioning from a transient causal state to a 
recurrent causal state on seeing $1$.}.  By Eq.~\eqref{transitionrev}, we have
\begin{align*}
\Pr\bigl(1, \nu(100) | \nu(00)\bigr)
  & = \Pr(1|00)  \\
  & = \nu(00) \widetilde{T}^{(1)} \one  \\
  & = \frac{1-p}{1+p+q-pq} ~.
\end{align*}

After constructing the mixed-state presentation, one calculates the stationary
state distribution. The causal states which have $\Pr(\PastCausalState) > 0$
are the recurrent causal states. These are $\PastCausalStateSet=\{D, E, F\}$:
\begin{align*}
D &= \nu(1001) =
  \bordermatrix{
     & A & B & C \cr
     & 0 & 0 & 1
  }\\
E &= \nu(100) =
  \bordermatrix{
     & A & B & C \cr
     & 1 & 0 & 0
  }\\
F &= \nu(10) =
  \bordermatrix{
     & A & B & C \cr
     & 0 & \frac{p}{p + q(1-p)} & \frac{q(1-p)}{p + q(1-p)}
  } ~.
\end{align*}
These mixed states give $\Pr(\FutureCausalState|\PastCausalState)$
which, when combined with $\Pr(\FutureCausalState)$, allows us to calculate:
\begin{equation*}
\EE = I[\PastCausalState; \FutureCausalState] = \FutureCmu - \FuturePC
\end{equation*}
with
\begin{equation*}
    \FutureCmu = 1 + \frac{H(p)}{2} 
\end{equation*}
and
\begin{equation*}
  \FuturePC = \frac{p+q(1-p)}{2} H\left( \frac{p}{p+q(1-p)} \right) ~,
\end{equation*}
where $H(\cdot)$ is the binary entropy function.

\section{Examples}

With the calculational procedure laid out, we now analyze the information
processing properties of several examples---two of which are familiar from 
symbolic dynamics.

\subsection{Even Process}

The Even Process is a stochastic generalization of the Even System: the
canonical example of a \emph{sofic subshift}---a symbolic dynamical system
that cannot be expressed as a subshift of finite type~\cite{Weis73,Crut01a}.
Although it has only two recurrent causal states, the Even Process cannot be
expressed as any finite Markov chain over measurement sequences. Somewhat
surprisingly, it turns out to be quite simple in terms of the properties we
are addressing. As we will now show, the mapping between forward and
reverse causal states is one-to-one and so $\PC = 0$. All of its internal
state information is present in measurements; we call it an \emph{explicit},
or \emph{non-cryptic} process.

\begin{figure}[th]
\centering
\includegraphics[scale=\figscale]{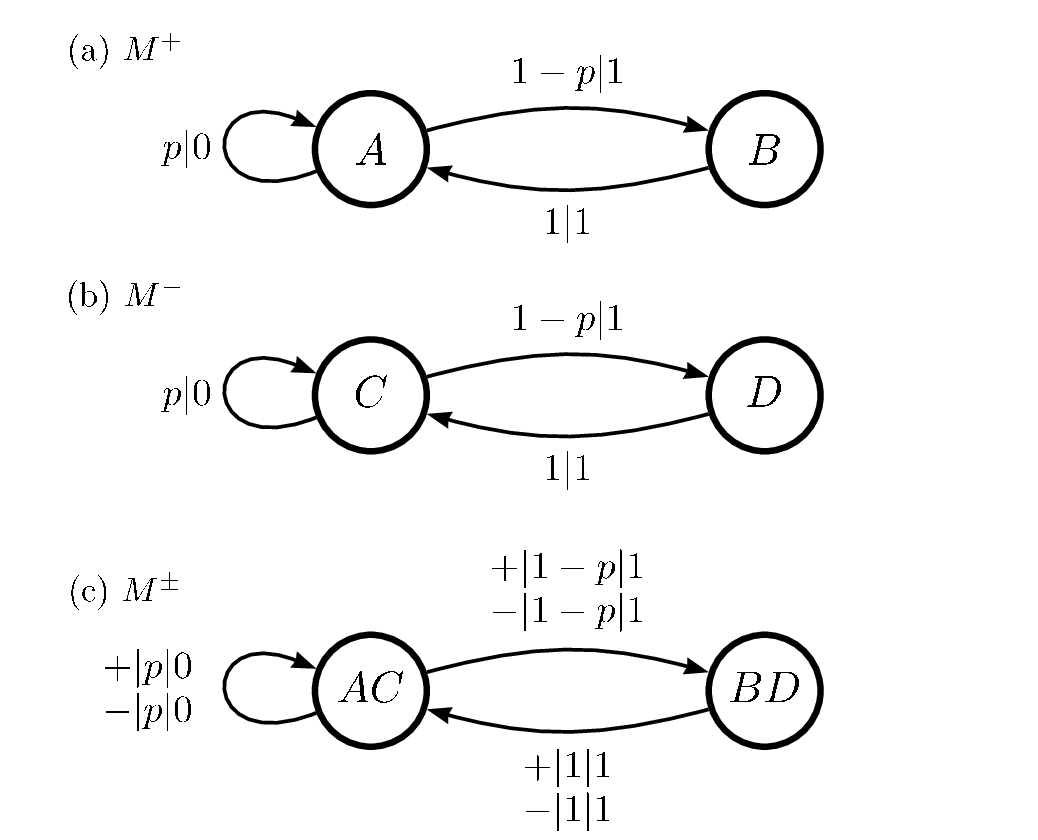}
\caption{
  Forward and reverse \eMs\ for the Even Process: (a) $\FutureEM$ and
  (b) $\PastEM$. (c) The bidirectional machine $\BiEM$.
  Edge labels are prefixed by the scan direction $\{-,+\}$.
  }
\label{fig:EvenProcess}
\end{figure}

Its forward \eM\ has two recurrent causal states
$\FutureCausalStateSet = \{ A, B \}$ and transition matrices~\cite{Crut01a}:
\begin{align*}
T^{(0)} &=
  \bordermatrix{%
      & A & B \cr
    A & p & 0 \cr
    B & 0   & 0
  }
  \mathrm{~and}\\
T^{(1)} &=
  \bordermatrix{%
      & A & B \cr
    A & 0 & 1-p \cr
    B & 1 & 0
  } ~.
\end{align*}
Figure \ref{fig:EvenProcess}(a) gives $\FutureEM$, while
\ref{fig:EvenProcess}(b) gives $\PastEM$. We see that the \eMs\ are the same
and so the Even Process is causally reversible ($\CI = 0$). Note
that $\widetilde{M}^+$ is unifilar.

We can give general expressions for the information processing invariants 
as a function of the probability $p = \Pr(0|A)$ of the self-loop.
A simple calculation shows that
\begin{align*}
\Pr(\FutureCausalState) &=
  \bordermatrix{
      & A & B \cr
      & \tfrac{1}{2-p} & \tfrac{1-p}{2-p}
  }
  \mathrm{~and}\\
\Pr(\PastCausalState) &=
  \bordermatrix{
      & C & D\cr
      & \tfrac{1}{2-p} & \tfrac{1-p}{2-p}
  } ~.
\end{align*}
And so, $\Cmu = H \left( 1/(2-p) \right)$ and $\hmu = H(p)/(2-p)$.
Since $\PC = 0$ for all $p$, we have $\EE = \Cmu$.

Now, let's analyze its bidirectional machine, which is shown in 
\figref{fig:EvenProcess}(c). The reverse and forward maps are given by:
\begin{align*}
\Pr(\FutureCausalState|\PastCausalState) &=
  \bordermatrix{%
      & A & B \cr
    C & 1 & 0 \cr
    D & 0 & 1
  }
  \mathrm{~and}\\
\Pr(\PastCausalState|\FutureCausalState) &=
  \bordermatrix{%
      & C & D \cr
    A & 1 & 0 \cr
    B & 0 & 1
  } ~.
\end{align*}
From which one calculates that
$\Pr(\BiCausalState) = \Pr(AC,BD) = \left(2/3,1/3\right)$ for $p=1/2$. 
This and the switching maps above give 
    $\BiCmu = H[\BiCausalState] = H(2/3) \approx 0.9183$
bits and $\EE = I[\FutureCausalState;\PastCausalState] \approx 0.9183$ bits.

Direct inspection of $\FutureEM$ and $\PastEM$ shows that both \eMs\ are
reverse unifilar. And this is reflected in the fact that 
$\FutureCmu = \PastCmu = \EE$; verifying a proposition of \refcite{Crut08d}.

\begin{figure}[th]
\centering
\resizebox{3.75in}{!}{\includegraphics{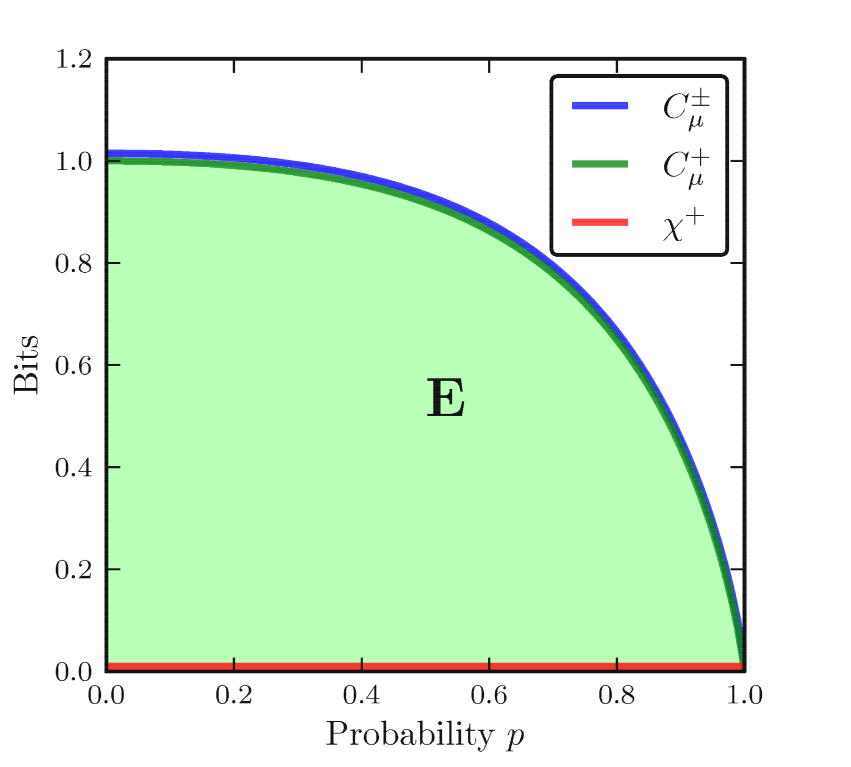}}
\caption{The Even Process's information processing properties---$\BiCmu$,
  $\FutureCmu$, and $\PC^+$---as its self-loop probability $p$ varies.
  The colored area bounded by the curves show the magnitude of $\EE$.
  }
\label{fig:EP_info}
\end{figure}

Without going into details to be reported elsewhere, the Even Process is also
notable since it is difficult to empirically estimate its $\EE$.
(The convergence as a function of the number of measurements is extremely slow.)
Viewed in terms of the quantities $\FutureCmu$, $\PastCmu$, $\chi^+$, $\chi^-$,
and $\Xi$, though, it is quite simple. This illustrates one strength of the
time-symmetric analysis. The latter's new and independent set of informational
measures lead one to explore new regions of process space (see Fig.
\ref{fig:EP_info}) and to ask structural questions
not previously capable of being asked (or answered, for that matter). To see
exactly why the Even Process is so simple, let's look at its causal states.

Its histories can be divided into two classes: those that end with an even
number of $1$s and those that end with an odd number of $1$s. Similarly, its
futures divide into two classes: those that begin with an even number of $1$s
and those that begin with an odd number of $1$s. The analysis here shows that
these classes are causal states $A$, $B$, $C$, and $D$, respectively; see
Fig. \ref{fig:EvenProcess}.

Beginning with a bi-infinite string, wherever we choose to split it into
$(\Past, \Future)$, we can be in one of only two situations: either
$(A,C)$ or $(B,D)$,
where $A$ ($C$) ends (begins) with an even number of 1s, and $B$ ($D$) ends
(begins) with an odd number of 1s. This one-to-one correspondence
simultaneously implies causal reversibility ($\Xi = 0$) and explicitness
($\PC = 0$). Thinking in terms of the bidirectional
machine, we can predict and retrodict, changing direction as often as we like 
and forever maintain optimal predictability and retrodictability. Since we can 
switch directions with no loss of information, there is no asymmetry in the
loss; this reflects the process's causal reversibility.

Plotting $\FutureCmu$, $\BiCmu$, and $\FuturePC$, \figref{fig:EP_info} rather
directly illustrates these properties and shows that they are maintained
across the entire process family as the self-loop probability $p$ is varied.

\subsection{Golden Mean Process}

The Golden Mean Process generates all binary sequences except for those with
two contiguous $0$s. Like the Even Process, it has two recurrent causal states.
Unlike the Even Process, its support is a subshift of finite type; describable by a
chain over three Markov states that correspond to the length-$2$ words $01$,
$10$, and $11$. Nominally, it is considered to be a very simple process.
However, it reveals several surprising subtleties. $\FutureEM$ and $\PastEM$
are the same \eM---it is causally reversible ($\CI=0$). However, $\BiEM$ has three
states and the forward and reverse state maps are no longer the identity. Thus,
$\PC > 0$ and the Golden Mean Process is cryptic and so hides much of its
state information from an observer.

\begin{figure}[th]
\centering
\includegraphics[scale=\figscale]{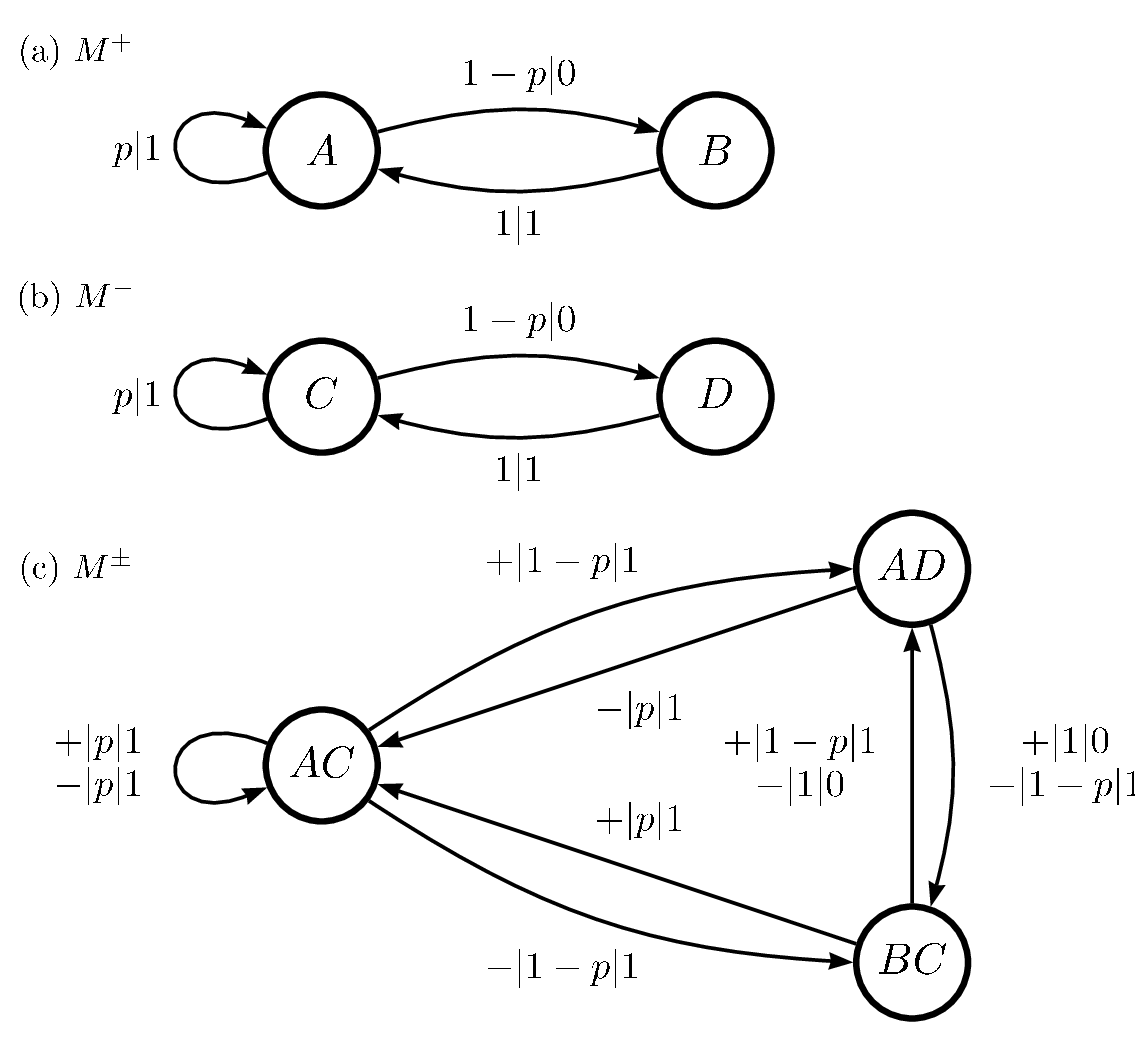}
\caption{
  Forward and reverse \eMs\ for the Golden Mean Process: (a) $\FutureEM$ and
  (b) $\PastEM$. (c) The bidirectional machine $\BiEM$.
  }
\label{fig:GMP}
\end{figure}

Its forward \eM\ has two recurrent causal states
$\FutureCausalStateSet = \{ A, B \}$ and transition matrices~\cite{Crut01a}:
\begin{align*}
T^{(0)} &=
  \bordermatrix{%
      & A & B \cr
    A & 0 & 1-p \cr
    B & 0   & 0
  }
\end{align*}
and
\begin{align*}
T^{(1)} &=
  \bordermatrix{%
      & A   & B \cr
    A & p & 0 \cr
    B & 1 & 0
  } ~.
\end{align*}
Figure \ref{fig:GMP}(a) gives $\FutureEM$, while (b) gives $\PastEM$. We
see that the \eMs\ are the same and so the Golden Mean Process is causally
reversible ($\CI = 0$).

Again, we can give general expressions for the information processing invariants as a function
of the probability $p = \Pr(1|A)$ of the self-loop. The state-to-state
transition matrix is the same as that for the Even Process and we also have
the same causal state probabilities. Thus, we have
$\Cmu = H \left( 1/(2-p) \right)$ and $\hmu = H(p)/(2-p)$ again, 
just as for the Even Process above. 
Indeed, a quick comparison of the state-transition diagrams does not
reveal any overt difference with the Even Process's \eMs.

\begin{figure}[th]
\centering
\resizebox{3.75in}{!}{\includegraphics{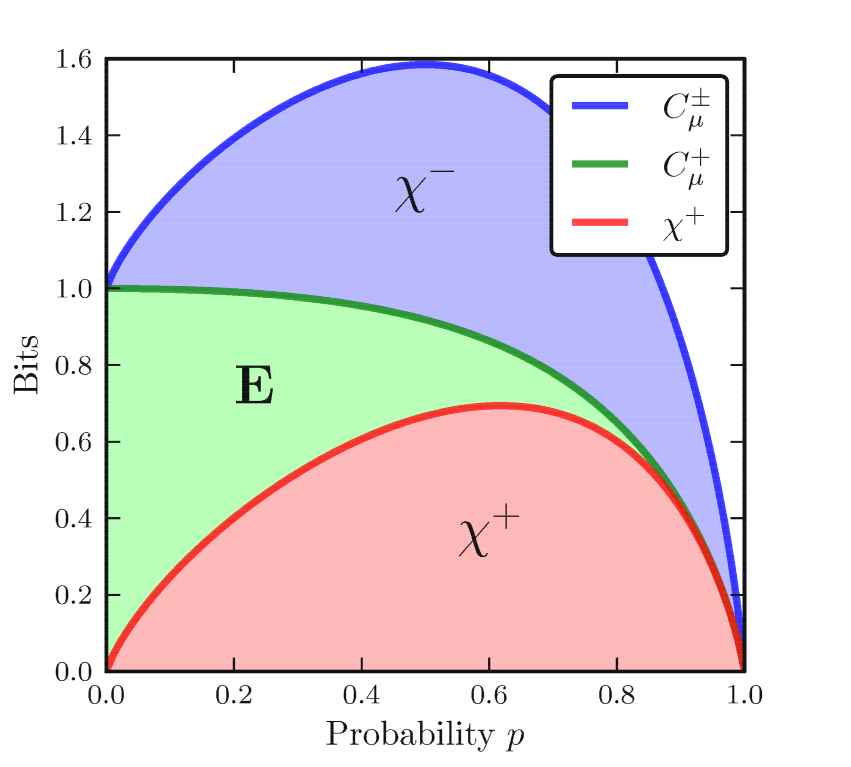}}
\caption{The Golden Mean Process's information processing
  invariants---$\BiCmu$, $\FutureCmu$, and $\PC^+$---as its self-loop
  probability $p$ varies. Colored areas bounded by the curves give the
  magnitude at each $p$ of $\PC^-$, $\EE$, and $\PC^+$.
  }
\label{fig:GMP_info}
\end{figure}

However, since $\PC \neq 0$ for $p \in (0,1)$ and since the process is also a
one-dimensional spin chain, we have \mbox{$\EE = \Cmu - R \hmu$ with $R = 1$}.
(Recall Eq. (\ref{SpinEEandCmu}).) Thus,
\begin{equation}
\EE = H \left( \frac{1}{2-p} \right) - \frac{H(p)}{2-p} ~.
\end{equation}
Putting these closed-form expressions together gives us a
graphical view of how the various information invariants change as
the process's parameter is varied. This is shown in Fig. \ref{fig:GMP_info}.

In contrast to the Even Process, the excess entropy is substantially less than
the statistical complexities, the signature of a cryptic process:
$\PC = H(p)/(2-p)$.

The origin of its crypticity is found by analyzing the bidirectional machine,
which is shown in \figref{fig:GMP}(c). The reverse and forward maps are
given by:
\begin{align*}
\Pr(\FutureCausalState|\PastCausalState) &=
  \bordermatrix{%
      & A   & B \cr
    C & p & 1-p \cr
    D & 1   & 0
  }
  \mathrm{~and}\\
\Pr(\PastCausalState|\FutureCausalState) &=
  \bordermatrix{%
      & C   & D \cr
    A & p & 1-p \cr
    B & 1 & 0
  } ~.
\end{align*}
From $\BiEM$, one can calculate the stationary distribution over the
bidirectional causal states:
$\Pr(\BiCausalState) = \Pr(AC,AD,BC) = \left(p,1-p,1-p\right)/(2-p)$. For
$p=1/2$, we obtain $\BiCmu = H[\BiCausalState] = \log_2 3 \approx 1.5850$
bits, but an $\EE = I[\FutureCausalState;\PastCausalState] = 0.2516$ bits. Thus,
$\EE$ is substantially less that the $\Cmu$s, a cryptic process:
$\PC \approx 1.3334$ bits.

The Golden Mean Process is a perfect complement to the Even Process.
Previously, it was viewed as a simple process for many reasons: It is based on
a subshift of finite type and order-$1$ Markov, the causal-state process is
\emph{itself} a Golden Mean Process, it is microscopically reversible, and
$\EE$ was exactly calculable (even before the introduction of the methods here).
However, the preceding analysis shows that the Golden Mean Process displays a
new feature that the Even Process does not---crypticity.

We can gain an intuitive understanding of this by thinking about classes
of histories and futures. In this case, a bi-infinite string can be split in
three ways $(\Past, \Future)$: $(A,C)$, $(A,D)$, or $(B,C)$,
where $A$ ($C$) is any past (future) that ends (begins) with a 0 and $B$
($D$) is any past (future) that ends (begins) with a 1. In terms of the
bidirectional machine, there is a cost associated with changing direction. 
It is the \emph{mixing} among the causal states above that is responsible
for this cost. Further, this cost is symmetric because of the microscopic reversibility. Switching
from prediction to retrodiction causes a loss of $\chi^+$ bits of memory and
a generation of $\chi^-$ bits of uncertainty.

Each complete round-trip state switch (e.g., forward-backward-forward)
leads to a geometric reduction in
state knowledge of $\EE^2/(\FutureCmu \PastCmu)$. One can characterize
this information loss with a half-life---the number of complete switches
required to reduce state knowledge to half of its initial value. 

Figure \ref{fig:GMP_info} shows that these properties are maintained across
the entire Golden Mean Process family, except at extremes. When $p = 0$, it
degenerates to a simple period-$2$ process, with
$\EE = \FutureCmu = \PastCmu = \BiCmu = 1$ bit of memory. When $p = 1$, it
is even simpler, the period-$1$ process, with no memory. As it approaches this
extreme, $\EE$ vanishes rapidly, leaving processes with internal state memory
dominated by crypticity: $\BiCmu \approx \FuturePC + \PastPC$.

\subsection{Random Insertion Process}

Our final example is chosen to illustrate what appears to be the typical
case---a cryptic, causally irreversible process. This is the random insertion
process (RIP) which generates a random bit with bias $p$. If that bit is a
$1$, then it outputs another $1$. If the random bit is a $0$, however, it
inserts another random bit with bias $q$, followed by a $1$.

\begin{figure}[th]
\centering
\includegraphics[scale=\figscale]{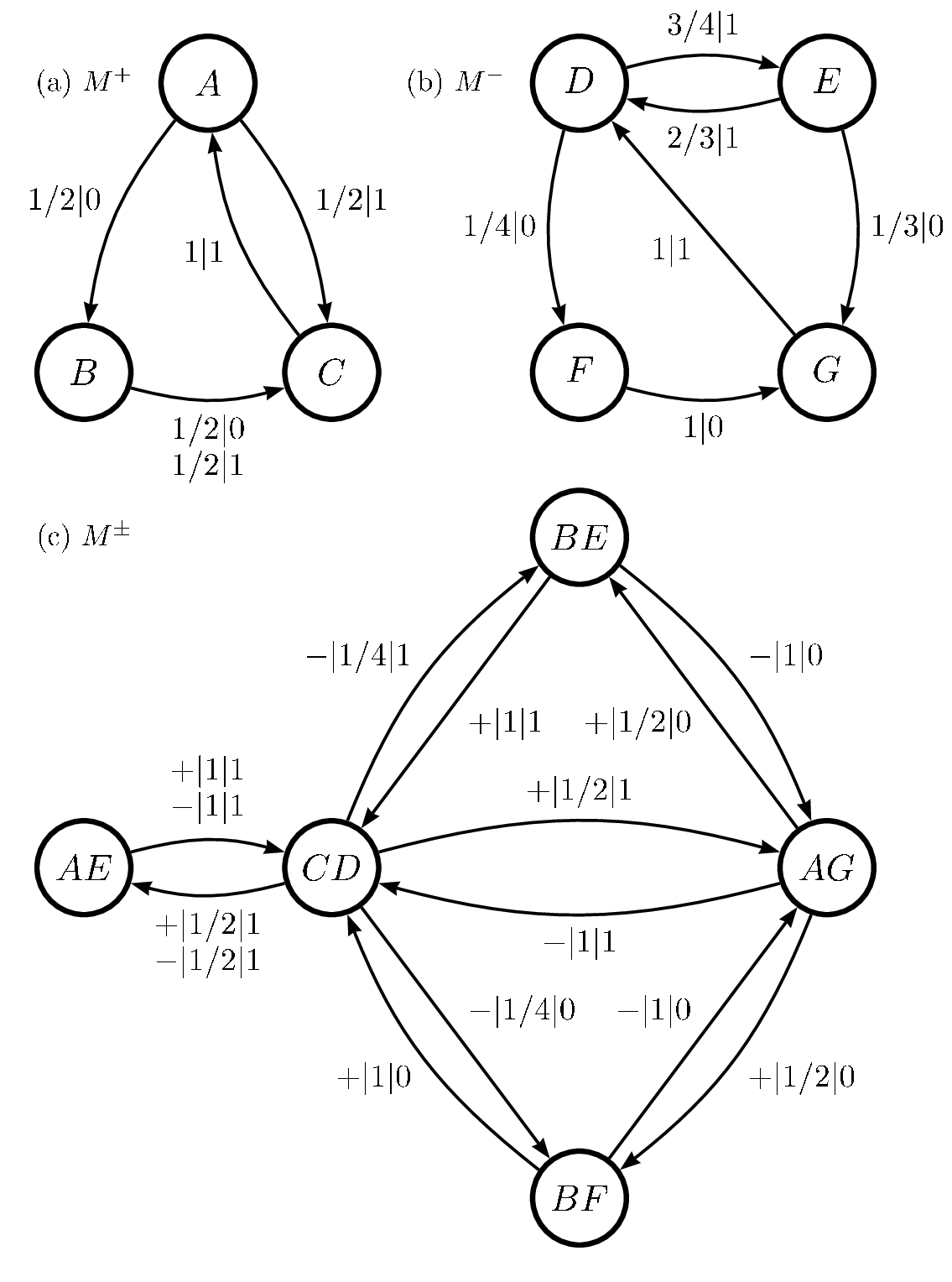}
\caption{
  Forward and reverse \eMs\ for the RIP with $p=q=1/2$: (a) $\FutureEM$ and
  (b) $\PastEM$. (c) The bidirectional machine $\BiEM$ also for $p = q = 1/2$.
  (Reprinted with permission from \protect\refcite{Crut08a}.)
  }
\label{fig:RIP}
\end{figure}

Its forward \eM\ has three recurrent causal states
$\FutureCausalStateSet = \{ A, B, C \}$ and transition matrices:
\begin{align*}
T^{(0)} &=
  \bordermatrix{%
      & A & B   & C \cr
    A & 0 & p & 0 \cr
    B & 0 & 0   & q \cr
    C & 0 & 0   & 0
  }
  \mathrm{~and}\\
T^{(1)} &=
  \bordermatrix{%
      & A & B   & C \cr
    A & 0 & 0   & 1-p \cr
    B & 0 & 0   & 1-q \cr
    C & 1 & 0   & 0
  } ~.
\end{align*}
Figure \ref{fig:RIP}(b) shows $\PastEM$ which has four recurrent causal states
$\PastCausalStateSet = \{ D, E, F, G \}$. We see that the \eMs\ are not the
same and so the RIP is causally irreversible. A direct calculation gives:
\begin{align*}
\Pr(\FutureCausalState) &=
  \bordermatrix{
      & A & B  & C \cr
      & \tfrac{1}{p+2} & \tfrac{p}{p+2} & \tfrac{1}{p+2}
  }
  \mathrm{~and}\\
\Pr(\PastCausalState) &=
  \bordermatrix{
      & D   & E    & F    & G \cr
      & \tfrac{1}{p+2} & \tfrac{1-pq}{p+2} & \tfrac{pq}{p+2} & \tfrac{p}{p+2}
  } ~.
\end{align*}
If $p = q = 1/2$, for example, these give us $\FutureCmu \approx 1.5219$ bits,
$\PastCmu \approx 1.8464$ bits, and $\hmu = 3/5$ bits per measurement.
The causal irreversibility is $\CI \approx 0.3245$ bits.

\begin{figure*}[th]
\centering
\resizebox{5.75in}{!}{\includegraphics{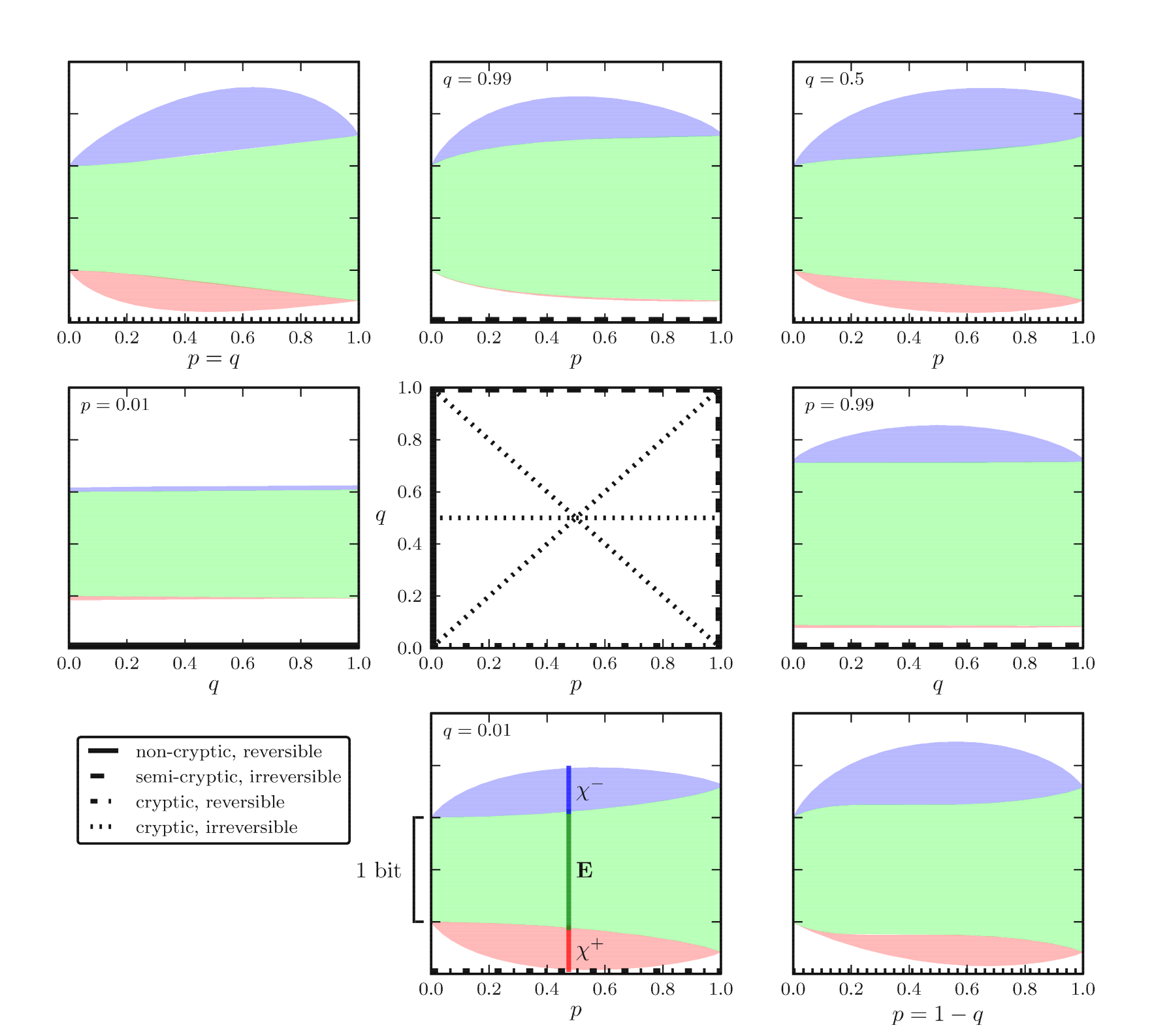}}
\caption{The Random Insertion Process's information processing
  invariants as its two
  probability parameters $p$ and $q$ vary. The central square
  shows the $(p,q)$ parameter space, with solid and dashed lines
  indicating the paths in parameter space for each of the other
  information versus parameter plots. The latter's vertical axes
  are scaled so that two tick marks measure $1$ bit of information.
  The inset legend indicates the class of process illustrated by
  the paths. Colored areas give the magnitude of
  $\PC^-$, $\EE$, and $\PC^+$.
  }
\label{fig:RIP_info}
\end{figure*}

Let's analyze the RIP bidirectional machine, which is shown in 
\figref{fig:RIP}(c) for $p = q = 1/2$. The reverse and forward maps
are given by:
\begin{align*}
\Pr(\FutureCausalState|\PastCausalState) &=
  \bordermatrix{%
      & A   & B   & C \cr
    D & 0   & 0   & 1 \cr
    E & 2/3 & 1/3 & 0 \cr
    F & 0   & 1   & 0 \cr
    G & 1   & 0   & 0 \cr
  }
  \mathrm{~and}\\
\Pr(\PastCausalState|\FutureCausalState) &=
  \bordermatrix{%
      & D  & E   & F   & G \cr
    A & 0  & 1/2 & 0   & 1/2 \cr
    B & 0  & 1/2 & 1/2 & 0 \cr
    C & 1  & 0   & 0   & 0 \cr
  } ~.
\end{align*}
Or, for general $p$ and $q$, we have
\begin{equation*}
\Pr(\FutureCausalState, \PastCausalState) = \frac{1}{(p+2)}
\bordermatrix{
   & D & E & F & G \cr
 A & 0 & 1-p & 0 & p \cr
 B & 0 & p(1-q) & pq & 0 \cr
 C & 1 & 0 & 0 & 0
} .
\end{equation*}
By way of demonstrating the exact analysis now possible, $\EE$'s closed-form
expression for the RIP family is
\begin{align*}
\EE = \log_2 (p+2) - \frac{p \log_2 p }{p + 2} -
\frac{1-pq}{p + 2} H\left( \frac{1-p}{1-pq}\right) ~.
\end{align*}
The first two terms on the RHS are $\FutureCmu$ and the last is $\FuturePC$.

Setting $p = q = 1/2$, one calculates that
$\Pr(\BiCausalState) = \Pr(AE,AG,BE,BF,CD) = \left(1/5,1/5,1/10,1/10,2/5\right)$.
This and the joint distribution give
$\BiCmu = H[\BiCausalState] \approx 2.1219$ bits,
but an $\EE = I[\FutureCausalState;\PastCausalState] = 1.2464$ bits. That is,
the excess entropy (the apparent information) is substantially less than the
statistical complexities (stored information)---a moderately cryptic process:
$\PC \approx 0.8755$ bits.

Figure \ref{fig:RIP_info} shows how the RIP's informational character varies
along one-dimensional paths in its parameter space: $(p,q) \in [0,1]^2$.
The four extreme-$p$ and -$q$ paths illustrate that the RIP borders on
(i) noncryptic, reversible processes (solid line), (ii) semi-cryptic,
irreversible processes (long dash), (iii) cryptic, reversible processes
(short dash), and (iv) cryptic, irreversible processes (very short dash).
The horizontal path ($q = 0.5$) and two diagonal paths ($p = q$
and $p = 1 - q$) show the typical cases within the parameter space
of cryptic, irreversible processes.

\section{Conclusions}

Casting stochastic dynamical systems in a time-agnostic framework revealed a
landscape that quickly led one away from familiar entrances, along new and
unfamiliar pathways. Old informational quantities were put in a new light,
new relationships among them appeared, and explicit calculation methods
became available. The most unexpected appearances, though, were the new
informational invariants that emerged and captured novel properties of
general processes.

Excess entropy, a familiar quantity in a long-applied family of mutual
informations, is often estimated
\cite{Fras90b,Casd91a,Spro03a,Kant06a,Arno96,Crut97a,Feld98b,Tono94a,Bial00a,Ebel94c,Debo08a}
and is broadly considered an important information measure for organization in
complex systems. The exact analysis afforded by our time-agnostic framework
gave an important calibration in our studies. Specifically, it showed how
difficult accurate estimates of the excess entropy can be. While we intend to
report on this in some detail elsewhere, suffice it to say that the convergence
of empirical estimates of $\EE$, in even very benign (and low statistical
complexity)
cases, can be so slow as to make estimation computationally intractable.
This problem would never have been clear without the closed-form expressions.
It, with nothing else said, calls into doubt many of the reported uses
and estimations of excess entropy and related mutual information measures.

Fortunately, we now have access to the analytic calculation of the excess
entropy from the \eM. Note that the latter is no more difficult to estimate
than, say, estimating the entropy rate of an information source. (Both are
dominated by obtaining accurate estimates of a process's sequence distribution.)
Notably,
the calculation relied on connecting prediction and retrodiction, which we
accomplished via the composition of the time-reversal operation on \eMs\ and
the mixed-state-presentation algorithm. As the analyses of the various example
processes illustrated, the technique yields closed-form expressions for $\EE$.
More generally, though, the explicit relationship between a process's \eM\ and
its excess entropy clearly demonstrates why the statistical complexity, and
not the excess entropy, is the information stored in the present.

In addition to the analytical advantage of having $\EE$ in hand, we learned a
pointed lesson about the difference between prediction (reflected
in $\EE$) and modeling (reflected in $\Cmu$). In
particular, a system's causal representation yields more direct access to
fundamental invariants than others---such as, histograms of word counts or
general hidden Markov models. The differences between prediction and modeling
unearthed new informational quantities---crypticity and causal irreversibility.

Crypticity describes the amount of stored state information that is not shared
in the measurement sequence. One might think of this as ``wasted'' information, 
although the minimality of the \eM\ suggests that this waste is necessary---that
is, an intrinsic property of the process. Possibly we could better think of
this as modeling overhead.

When analyzing time symmetry, one can use notions such as microscopic
reversibility or, more broadly, reversible support. We introduced the
yet-broader notion of causal irreversibility $\Xi$. It has the advantage of
being scalar rather than Boolean and so has something to say quantitatively
about all processes. Also, it derives naturally from its simple relationship to
$\EE$ and $\PC$. In this light, microscopic reversibility appears to be too
strong a criterion, missing important structural properties.

The time-agnostic perspective hinged on expanding the space of representations.
First, we described parallel predictive and retrodictive causal models joined
by the switching maps. We then introduced a bidirectional machine that compressed
$\FutureCmu$ and $\PastCmu$ into $\BiCmu$. The associated joint causal-state
space allowed us to make rather nonintuitive statements about prediction
(retrodiction) conditioned on these joint states. The operational meaning
of the bidirectional machine certainly warrants further attention. It also
seems likely that its nonunifilarity has not yet been fully appreciated. One
might wish to consider, for example, a unifilar representation of it.
Somewhat hopefully, we end by noting that the bidirectional machine suggests
an extension of \eM\ analysis beyond one-dimensional processes.

\section*{Acknowledgments}

Chris Ellison was partially supported on a GAANN fellowship. The Network
Dynamics Program funded by Intel Corporation also partially supported this
work.

\appendix

\section{Appendix: The Mixed-State Presentation is Sufficient to
Calculate the Switching Maps}

While we conjecture that the mixed-state operation $\MSP(\widetilde{M}^+)$
yields an $\eM$, this remains an open problem. Our conjecture, however, is
based on a rather large number of test cases in which it is an \eM.
Fortunately for our present needs, we can show that
$\MSP(\widetilde{M}^+)$ is sufficient for
calculating the conditional probability distribution
$\Prob(\FutureCausalState | \PastCausalState)$.

For a moment, ignore the details of forward and reverse machines and simply
consider machines $A$ and $B$ such that $\MSP(A) = B$ where neither $A$ nor $B$ is
necessarily an $\eM$. We would like to learn the conditional probability
distribution $\Prob(\AlternateState_A | \AlternateState_B)$, where
$\AlternateState_A$ and $\AlternateState_B$ are $A$'s and $B$'s states,
respectively.

\begin{Prop}
$B$'s states are mixed states of $A$.
\end{Prop}

\begin{ProProp}
We use the mixed-state presentation algorithm to form states based on the
transition matrices of $A$. If a state $\AlternateState_B$ is induced by a
word $w$, then:
\begin{align*}
\AlternateState_B = \frac{\pi_A T_A^{\omega}}{\pi_A T_A^{w} \one} ~.
\qed
\end{align*}
\end{ProProp}

We now show that $B$ is deterministic.

\begin{Prop}
$H[\AlternateState' | \AlternateState, \MeasSymbol] = 0$ for machine $B$.
\end{Prop}

\begin{ProProp}
Although any given state in $B$ will generally be a distribution over states
in $A$, each of these distributions \emph{defines} a state of $B$. The
particular state of $B$ (or distribution over states in $A$),
$\AlternateState'$, that follows $\AlternateState$ and $\MeasSymbol$ can be
written:
\begin{align*}
\AlternateState'_B
  = \frac{\pi_A T_A^{\omega} T^{\MeasSymbol}}
    {\pi_A T_A^{\omega} T^{\MeasSymbol} \eta} ~.
\end{align*}
So, by construction, $B$ is deterministic.
\qed
\end{ProProp}

Moreover, $\AlternateState_B$ is a refinement of $\CausalStateSet_B$.

\begin{Prop}
Two pasts that induce the same state in $B$ must be pasts in the same
causal state of $B$'s \eM.
\end{Prop}

\begin{ProProp}
The future probability distribution given a word is exactly the future
probability distribution given the mixed state induced by that word:
\begin{align*}
\Prob(\Future | \omega)
  & = \frac{\pi T^{\omega} T^{\Future}}
  	{\pi T^{\omega} T^{\Future} \eta}\\
\Prob(\Future | \mu(\omega))
  & = \frac{\frac{\pi T^{\omega}}{\pi T^{\omega} \eta} T^{\Future}}
  	{\frac{\pi T^{\omega} T^{\Future} \eta}{\pi T^{\omega} \eta}}
  = \frac{\pi T^{\omega} T^{\Future}}{\pi T^{\omega} T^{\Future} \eta}
\end{align*}
Therefore, if two words induce the same mixed state, the future probability
distribution conditioned on those words are the same. This means that those
words are causally equivalent and thus in the same causal state.
\qed
\end{ProProp}

Now we show how, even in this very generic case, we can calculate the relevant
conditional probability distribution.

The mixed-state construction of $B$ implicitly has given us
$\Prob(\AlternateState_A | \AlternateState_B)$, which we can use to
find $\Prob(\AlternateState_A | \CausalState_B)$, our goal:
\begin{align*}
\Prob(\AlternateState_A | \CausalState_B)
  &= \sum_{\AlternateState_B}
  	\Prob(\AlternateState_A | \CausalState_B, \AlternateState_B)
	\Prob(\AlternateState_B | \CausalState_B) \\
  &= \sum_{\AlternateState_B}
  	\Prob(\AlternateState_A | \AlternateState_B)
	\Prob(\AlternateState_B | \CausalState_B) \\
  &= \sum_{\AlternateState_B}
  	\Prob(\AlternateState_A | \AlternateState_B)
	\Prob(\CausalState_B | \AlternateState_B)
	\frac{\Prob(\AlternateState_B)}{\Prob(\CausalState_B)} \\
  &= \sum_{\AlternateState_B}
  	\Prob(\AlternateState_A | \AlternateState_B)
	\delta_{\AlternateState_B \in \CausalState_B}
	\frac{\Prob(\AlternateState_B)}{\Prob(\CausalState_B)} \\
  &= \sum_{\AlternateState_B}
  	\Prob(\AlternateState_A | \AlternateState_B)
	\frac{\Prob(\AlternateState_B)}{\Prob(\CausalState_{\AlternateState_B})} ~.
\end{align*}
The second line follows since $\AlternateStateSet_B$ is a refinement of
$\CausalStateSet_B$. The third line is an application of Bayes Rule. The
fourth line follows again from the refinement. The final form reminds us that
$\CausalStateSet_B$ is not a free variable.

To sum up, we calculate the conditional distribution using this final form as
follows. The first factor is found by applying $\MSP$ to $A$. Granting
ourselves the ability to ascertain predictive equality among a finite set of
states $\AlternateStateSet_B$, we determine if
$\AlternateState_B \in \CausalState_B$ for each $\AlternateState_B$. Lastly,
we compute the stationary distribution over the states of $B$ and divide by
the stationary probability of the corresponding causal state.

In effect, this establishes a general method for computing the conditional
probability of states from the ``input'' machine given a state of the
``resultant'' machine. We can now recall the specific context of forward
and reverse \eMs\ and apply this technique to calculate $\EE$ in the case
where the resultant machine $\TR(\FutureEM)$ is not an $\eM$.

The input machine is the reversed $\eM$ $\TR(\FutureEM)$, whose states
$\widetilde{\CausalStateSet}{}^+$ are in one-to-one correspondence with
$\FutureCausalStateSet$. Thus, the previous result:
\begin{align*}
\Prob(\AlternateState_A | \CausalState_B)
  &= \sum_{\AlternateState_B}
  {\Prob(\AlternateState_A | \AlternateState_B)
  \frac{\Prob(\AlternateState_B)}{\Prob(\CausalState_{\AlternateState_B})}}
\end{align*}
now becomes:
\begin{align*}
\Prob(\CausalState_A | \CausalState_B)
  &= \sum_{\AlternateState_B}{\Prob(\CausalState_A | \AlternateState_B)
  \frac{\Prob(\AlternateState_B)}{\Prob(\CausalState_{\AlternateState_B})}}
\end{align*}
or, more specifically,
\begin{align*}
\Prob(\FutureCausalState | \PastCausalState)
  &= \sum_{\AlternateState_B}
  	\Prob(\FutureCausalState | \AlternateState_B)
	\frac{\Prob(\AlternateState_B)}
	{\Prob(\PastCausalState_{\AlternateState_B})} ~.
\end{align*}
From which we readily calculate $\EE$ using:
\begin{align*}
\EE &= I[\FutureCausalState; \PastCausalState]\\
    &= H[\FutureCausalState] - H[\FutureCausalState | \PastCausalState] ~.
\end{align*}

\bibliography{ref,chaos}

\end{document}